\documentclass[twocolumn,pra,superscriptaddress]{revtex4-2}
\pdfoutput=1
\usepackage[utf8]{inputenc}
\usepackage[T1]{fontenc}
\usepackage{url}
\usepackage{lipsum}
\usepackage{xcolor}
\usepackage{physics}
\usepackage{orcidlink}
\usepackage[final]{changes}
\usepackage{comment}
\usepackage{graphicx}
\usepackage{dcolumn}
\usepackage{bm}
\usepackage{hyperref}
\usepackage{verbatim}

\begin{document}

\newcommand{\QuICS}{Joint Center for Quantum Information and Computer Science, National Institute of Standards and Technology and University of Maryland, College Park, Maryland 20742, USA}
\newcommand{\JQI}{Joint Quantum Institute, National Institute of Standards and Technology and
  University of Maryland, College Park, Maryland 20742, USA}
\newcommand{\JQIalt}{Joint Quantum Institute, National Institute of Standards and Technology\\ and University of Maryland, College Park, Maryland 20742, USA}
\newcommand{\QUICSalt}{Joint Center for Quantum Information and Computer Science, National Institute of\\ Standards and Technology and University of Maryland, College Park, Maryland 20742, USA}

\author{Deniz Kurdak\,\orcidlink{0000-0003-4076-3013}}
\affiliation{\JQI}
\email{dkurdak@umd.edu}
\author{Patrick R. Banner\,\orcidlink{0009-0006-9957-4996}}
\affiliation{\JQI}
\author{Yaxin Li\,\orcidlink{0000-0001-8734-0136}}
\affiliation{\JQI}
\author{Sean~R.~Muleady\,\orcidlink{0000-0002-5005-3763}}
\affiliation{\JQI}
\affiliation{\QuICS}
\author{Alexey V. Gorshkov\,\orcidlink{0000-0003-0509-3421}}
\affiliation{\JQI}
\affiliation{\QuICS}
\author{S. L. Rolston\,\orcidlink{0000-0003-1671-4190}}
\affiliation{\JQI}
\author{J. V. Porto\,\orcidlink{0000-0002-6290-7535}}
\affiliation{\JQI}

\preprint{APS/123-QED}

\title{Enhancement of Rydberg Blockade via Microwave Dressing}

\date{\today}

\begin{abstract}
Experimental control over the strength and angular dependence of interactions between atoms is a key capability for advancing quantum technologies.
Here, we use microwave dressing to manipulate and enhance Rydberg-Rydberg interactions in an atomic ensemble. 
By varying the cloud length relative to the blockade radius and measuring the statistics of the light retrieved from the ensemble, we demonstrate a clear enhancement of the interaction strength due to microwave dressing. 
These results are successfully captured by a theoretical model that accounts for the excitation dynamics, atomic density distribution, and the phase-matched retrieval efficiency. Our approach offers a versatile platform for further engineering interactions by exploiting additional features of the microwave fields, such as polarization and detuning, opening pathways for new quantum control
strategies.
\end{abstract}

\maketitle

{\it Introduction }---
Cold atoms excited to Rydberg levels are a leading platform for exploring a wide variety of physics, including quantum computation \cite{Saffman.2010}, simulation 
\cite{Browaeys.2020}, sensing \cite{Fan.2015}, and optics \cite{Firstenberg.2016}. 
Their utility arises from the strong scaling of fundamental atomic properties with principal quantum number \cite{Gallagher.1994}, such as the polarizability and dipole-dipole (DD) interaction strength.
Rydberg blockade, originally proposed in Refs.~\cite{Jaksch.2000,Lukin.2000}, leverages the enhanced atom-atom interactions to suppress the simultaneous Rydberg excitation of two atoms spaced closer than the ``blockade radius.'' 

Tuning the strength and angular dependence of the DD interaction, as well as its scaling with atom separation, would enhance and expand the capabilities of interacting Rydberg systems.
For example, in the context of quantum computation, increasing the strength of interactions can enable faster two-qubit gates with higher fidelity, as well as many-qubit entangling gates \cite{Saffman.2005,Levine.2018}. 
In quantum simulation, the ability to finely control Rydberg interactions has enabled the implementation of a broad class of Hamiltonians \cite{Glaetzle.2015,Barredo.2015,Hollerith.2022,Scholl.2022,Steinert.2023,Signoles.2021,Zeiher.2016,Borish.2020,Geier.2021}. 
In ensemble-based experiments with Rydberg polaritons, enhancing interactions would increase the optical depth per blockade radius --- the key experimental parameter enabling quantum nonlinearities at the single-photon level \cite{Peyronel.2012} --- permitting improved single-photon transistors, photonic quantum gates, generation of non-classical states of light and symmetry-protected collisions between photons
\cite{Kanungo.2022,Ornelas-Huerta.2020, Yang.2022, Shi.2022, Ye.2023, Tiarks.2019, Tiarks.2014, Gorniaczyk.2014, Gorshkov.2011, Grass.2018,Gullans.2017, Paredes-Barato.2013, Thompson.2017}. 

There has been considerable work on increasing the interaction strength of Rydberg atoms \cite{Tiarks.2014, Walker.2008, Bohlouli-Zanjani.2007, Tretyakov.2014, Ryabtsev.2010, Ravets.2014,Reinhard.2008, Nipper.2012, Gorniaczyk.2016, Tanasittikosol.2011, Brekke.2012,Xu.2024}.
One approach is to reduce F{\"o}rster defects leading to increased van der Waals (vdW) interactions by choosing Rydberg states that have naturally low F{\"o}rster defects \cite{Tiarks.2014, Walker.2008} or by using perturbing fields to bring states into resonance \cite{Bohlouli-Zanjani.2007,Tretyakov.2014,Ryabtsev.2010,Ravets.2014,Reinhard.2008,Nipper.2012,Gorniaczyk.2016}.
Another approach is to use microwave dressing to couple opposite parity Rydberg levels, generating superposition states which have stronger first-order interactions \cite{Tanasittikosol.2011,Brekke.2012,Xu.2024}.
Microwave dressing of Rydberg states has also been proposed to \textit{decrease} interactions, allowing for the creation of asymmetric Rydberg gates \cite{Young.2021}, nullification of vdW interactions \cite{Shi.2017}, and realization of spin-charge separation of dark-state polaritons \cite{Shi.2016}, among other novel applications \cite{Sevincli.2014}.

\begin{figure*}[ht!]
\includegraphics[width=\textwidth]{"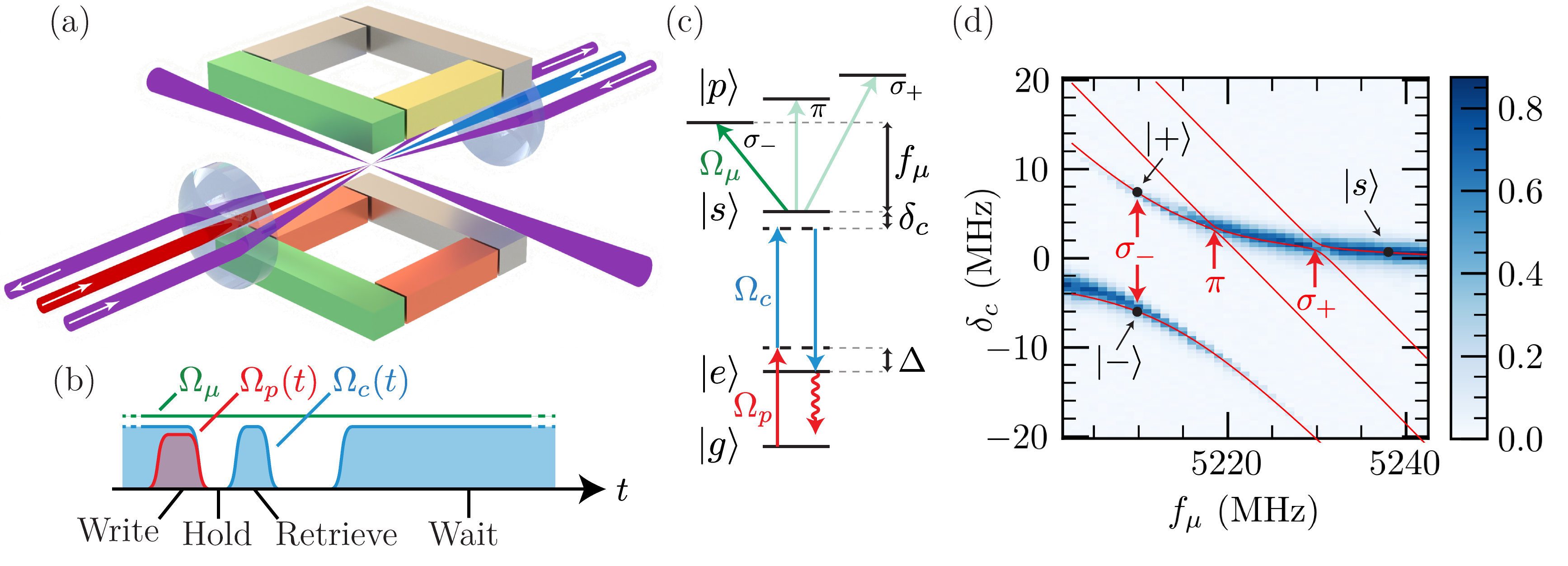"}
\caption{\label{fig:Fig1} (a) Experimental setup: Two in-vacuum lenses focus counter-propagating probe (red) and control (blue) beams onto an atomic cloud (not shown) trapped by a crossed-dipole trap (purple). Eight electrodes  cancel stray DC electric fields. The phases and amplitudes of microwaves applied to three sets of the electrodes (green, orange, and yellow) control the microwave polarization and amplitude at the cloud. (b) Schematic of the control, probe, and microwave fields during one cycle of the experiment. (c) Level diagram: 
A microwave field, with frequency $f_\mu$ and Rabi frequency $\Omega_\mu$, resonantly couples $\ket{s}$ and $\ket{p}$ creating the dressed states $\ket{\pm} = \frac{1}{\sqrt{2}}\left(\ket{s} \pm \ket{p}\right)$. The cloud is driven to the dressed or undressed Rydberg states, $\ket{-}$ or $\ket{s}$, from $\ket{g}$ by a two-photon process detuned $\Delta$ from the intermediate state $\ket{e}$. The excitations are extracted as photons by resonantly coupling $\ket{-}$ or $\ket{s}$ to $\ket{e}$. (d) 2D electromagnetically induced transparency (EIT) spectroscopy of dressed $88S_{1/2}$ vs. two-photon detuning $\delta_c$ and microwave frequency $f_\mu$. The color bar indicates probe transmission. The red lines are fits to the spectrum, where $\Omega_\mu/(2\pi) \approx 13$  MHz.  The avoided crossings associated with $\sigma_-$, $\pi$ and $\sigma_+$ polarizations are shown. Minimizing the $\sigma_+$ and $\pi$ components results in a $\sigma_-$ field purity $E_{\sigma_-}/\left|\mathbf{E}\right|=99.3(3)\%$ \cite{Supplemental}.}
\end{figure*}


In this work, we use our Rydberg-ensemble system \cite{Ornelas-Huerta.2020} to study the enhancement of Rydberg interactions due to microwave dressing. 
We generate spectroscopically resolved microwave-dressed eigenstates to which we write collective Rydberg excitations that are extracted phase coherently as photons.
The strength of the interactions determines the number of Rydberg excitations within the sample, and therefore the number of photons extracted per pulse.
To study the interactions, we thus measure the single-photon purity of the emitted light field as a function of the ensemble length.
Our experimental results show that the dressed eigenstates have significantly enhanced pair interactions.
To support our findings, we model the photon generation and retrieval process with a 1D pseudo-spin model, which uses Floquet calculations of the pair-state interactions and the experimentally measured atomic density profiles, yielding numerical results in good agreement with observations.
We show that, as a single-photon source, our system benefits from the enhanced interactions, which enable higher-purity single photons without compromising production efficiency.

Prior work utilizing microwaves in single-photon sources studied dephasing during the excitation storage time \cite{Bariani.2011,Bariani.2012,Maxwell.2013,Maxwell.2014,Fan.2023,Xu.2024}.
Our work instead examines the write stage, studying blockade physics in a regime of principal quantum numbers, sample sizes, and densities such that our ensemble only hosts a few Rydberg excitations, which allows us to coherently probe and dynamically model the system at the few-body level.
\newpage
{\it Experimental Setup }---
Our ensemble is a laser-cooled cloud of $^{87}$Rb atoms, held in a ``magic"-wavelength dipole trap \cite{Lampen.2018}.
The atoms are collected in a magneto-optical trap (MOT) and loaded into the dipole trap, where they are cooled to $\approx10$ $\mu$K via $\Lambda$-gray molasses~\cite{Rosi.2018} and optically pumped into the $\ket{g} \equiv \ket{5S_{1/2}, F=2, m_F = -2}$ state. 
We control the length of the cloud by adjusting the relative depths of two independent dipole traps (see Fig.~\ref{fig:Fig1}): a ``crossed'' trap composed of two nearly counter-propagating beams with orthogonal linear polarizations, intersecting the probe and control beams at angles of $\pm 11^\circ$, and a single-beam transverse trap, aligned perpendicular to the probe (see the Supplemental Material (SM) \cite{Supplemental}). After adjusting the trap depths to achieve the desired cloud length, the polarization of one of the crossed-trap beams is rotated to realize a 1D lattice, freezing the length of the cloud, and the transverse beam is turned off to avoid vector light shifts during Rydberg excitation.
The resulting cigar-shaped cloud has a transverse waist of 10~$\mu$m and a longitudinal root-mean-square (RMS) radius that can be adjusted between 20~$\mu$m and 50~$\mu$m.
The peak density of the ensemble increases with cloud length, ranging from $\approx 2\times10^{11}$ cm$^{-3}$ to $\approx 5\times10^{11}$ cm$^{-3}$.
The counter-propagating probe and control beams are focused to $\approx 3.3 \, \mu$m and $\approx 19 \,\mu$m waists, respectively.
On the time scales of the Rydberg pulse sequences, the positions of the atoms are effectively frozen, so the narrow waist of the probe defines the transverse extent of the ensemble that is addressed by the Rydberg lasers, resulting in an effectively 1D ensemble.

\begin{figure*}[t!]
\includegraphics[width=\textwidth]{"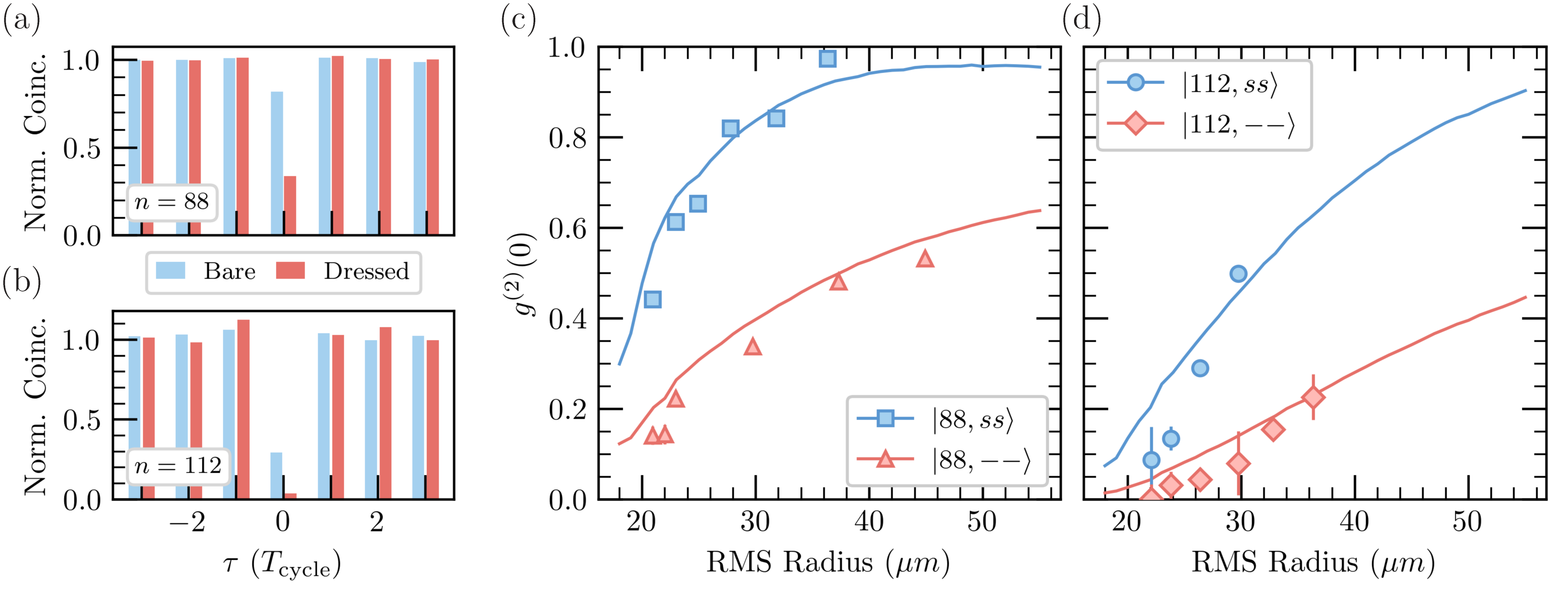"}
\caption{\label{fig:Fig2} (a,b) Suppression of $g^{(2)}(0)$ at $n=88$ and $n=112$. The bars indicate the normalized pulse-integrated coincidences between the two SPADs as a function of click separation $\tau$, in units of the cycle time, $T_\text{cycle}$. The coincidences have been gated and background subtracted \cite{Ornelas-Huerta.2020}. The blue and red bars, horizontally offset for visual clarity, show the statistics of light generated using bare and dressed eigenstates, respectively. (c,d) The variation of $g^{(2)}(0)$ as a function of RMS cloud radius. The data points in (c) and (d) correspond to the results gathered at $n=88$ and $n=112$, respectively. The error bars indicate the $\pm1 \sigma$ statistical uncertainties. The solid lines are numerical results of simulations, described below.} 
\end{figure*} 

{\it Single-photon generation }---
Following cloud shaping, the single-photon-generation cycle, consisting of a write, hold, and retrieve stage, was repeated $2.5 \times 10^4$ times per MOT cycle (see Fig.~\ref{fig:Fig1}(b)). 
During the write stage, the atoms in $\ket{g}$ were driven to the Rydberg manifold via a two-photon excitation through the intermediate state $\ket{e} \equiv \ket{5P_{3/2},F = 3, m_{F} = -3}$ with a detuning of $\Delta/(2\pi) = 50$ MHz (see Fig.~\ref{fig:Fig1}(c)). 
The control and probe Rabi frequencies driving the two-photon transitions were $\Omega_c/(2 \pi) \approx 6 \, \text{MHz}$ and $\Omega_p/(2 \pi) \approx 0.4 \, \text{MHz}$  for $n=112$ and $10\, \text{MHz}$ and $1.5\, \text{MHz}$ for $n=88$. 
The write-pulse duration, $\sim 200$ ns for $n=88$  and  $\sim 1 \, \mu$s for $n=112$, was chosen to maximize the photon generation efficiency for the given Rabi frequencies.
The  excitations were held for 200 ns while the control laser was brought on resonance, after which a final control pulse converted the phase-matched atomic excitations to photons propagating along the probe axis.
The write, hold, and retrieve sequence, sketched in Fig. 1(b), takes $\lesssim2 \, \mu$s, but the cycle was repeated only every 20 $\mu$s to allow for the decay of contaminant Rydberg excitations that are dark to the retrieval lasers \cite{Goldschmidt.2016, Ornelas-Huerta.2020}. 
The generated light was detected by a pair of single-photon avalanche detectors (SPADs).

To characterize the statistics of the retrieved light, we measure the coincidences between the two SPADs in a Hanbury Brown-Twiss (HBT) configuration, from which we estimate the pulse-integrated second-order correlation function $g^{(2)}(0) = \left< \hat{n} \left(\hat{n}-1 \right) \right>/\left<\hat{n}\right>^2$,
where $\hat{n}$ is the photon-number operator.
The deviation of $g^{(2)}(0)$  from zero indicates the breakdown of blockade, and we define the purity of the single photons as $1-g^{(2)}(0)$. 
HBT interferometers with click/no-click detectors are susceptible to bias at high photon rates, and would have caused errors on the order of a few percent in our highest-rate datasets \cite{Migdall.2013}. 
However, in our experiment, the photon pulses ($\sim 1\, \mu$s) are much longer than the instrument dead time ($\approx 25$~ns), and the SPADs are able to detect multiple photons per shot \cite{Banner.2024}.
The unbiased estimator for $g^{(2)}(0)$ including these multiple-click events is also immune to imperfections such as an unbalanced beam splitter \cite{Migdall.2013}.

{\it Microwave dressing }---
To modify the strength of the Rydberg-Rydberg interactions, we admix opposite-parity bare states, $\ket{s} \equiv \ket{nS_{1/2},m_J = -1/2}$ and $\ket{p}\equiv \ket{nP_{3/2},m_J = -3/2}$, with microwaves generated using the in-vacuum electrodes shown in Fig.~\ref{fig:Fig1}(a).
We use three sets of electrodes with independent phase and amplitude control to generate linearly independent microwave electric fields, see SM \cite{Supplemental}. 
The phases and amplitudes were chosen to produce pure $\sigma_-$ polarization at $\omega_\mu\approx 2\pi \times 5.21$ GHz and $\approx2\pi \times  2.47$ GHz to target the $nS_{1/2} - nP_{3/2}$ resonances at $n=88$ and $n=112$, respectively. The purified microwave polarizations were measured using EIT spectroscopy of avoided crossings (Fig. \ref{fig:Fig1}(d)) \cite{Robinson.2021}.

The atoms were excited to either the bare $\ket{s}$ states (microwaves off) or the microwave-dressed eigenstates.
In the latter case, the microwaves were on continuously and resonant with the $\ket{s}$ to $\ket{p}$ transition.
The resulting single-particle dressed states are $\ket{\pm} \equiv \frac{1}{\sqrt{2}} \left( \ket{s} \pm \ket{p} \right)$ with eigenvalues $E_\pm = \pm \hbar \Omega_\mu/2$, where $\Omega_\mu/(2 \pi) \approx6$ MHz is the microwave Rabi frequency.
For the results here, we excited to the lower-energy eigenstate $\ket{-}$, shown in Fig.~\ref{fig:Fig1}(d). 

We note that the dipole trap is close to magic for the $\ket{s}$ state. Since the admixture of the $\ket{p}$ state modifies the polarizability, the magic condition is not satisfied for the dressed states, leading to a reduction of the spin-wave coherence time from $\approx1.3\, \mu$s to $\approx0.8\, \mu$s. 

We measure $g^{(2)}(0)$ for bare and dressed Rydberg states at two principal quantum numbers, $n=88$ and $n=112$. 
Normalized coincidences are shown in Fig.~\ref{fig:Fig2}(a,b), where, for similar sized clouds ($\sim 27$~$\mu$m) at $n=88$ and $n=112$, dressing reduced the $g^{(2)}(0)$ from 0.82(1) to 0.34(1) and from 0.29(2) to 0.04(2), respectively. 
The number of atom pairs separated by distances greater than the blockade radius $r_b$ is modified by varying the length of the cloud (see SM~\cite{Supplemental}) to characterize the differences between the bare- and dressed-state interactions.
We show the results in Fig.~\ref{fig:Fig2}(c,d) for bare and dressed states at the two $n$.
For the bare $n=112$ state, we measure a rapid increase of $g^{(2)}(0)$ as the cloud size increases above the minimum RMS radius of 22 $\mu$m.
With dressing, we observe that $g^{(2)}(0)$ remains below 0.1 for RMS radii up to $\approx 30 \, \mu$m, indicating a significant increase of the blockade radius. 
We observe similar suppression of $g^{(2)}(0)$ at $n=88$, also showing that the dressing-induced interactions  at $n=88$ are comparable to the bare vdW interactions  at $n=112$, despite the bare vdW interaction strengths at a given distance differing by a factor of $\sim16$.

The increase in interaction strength can be understood by investigating the DD interaction of a pair of Rydberg atoms, given by \cite{Reinhard.2007}
\begin{align}
    \hat{V}_{dd}\left(\mathbf{r}\right) &= -\frac{1}{4 \pi \epsilon_0} \frac{1}{r^3} \sqrt{\frac{24 \pi}{5}} \sum_{\mu,\nu} \bra{1, \mu; 1, \nu}\ket{2,\mu+\nu}\\ \nonumber
    &\hspace{100pt}\times  Y^*_{2, \mu+\nu} \left(\theta, \phi\right) \, \hat{d}^{(1)}_\mu \hat{d}^{(2)}_\nu ,
\end{align}
where $\mu, \nu$ are the polarizations  summing over $\pm1$ and $0$ , $Y_{l,m}$ are the spherical harmonics, and $\hat{d}^{(1,2)}$ are the dipole operators for the two atoms. 
Due to selection rules of $\hat{d}$, a pair of atoms in bare states experience no first-order DD interaction, i.e. $\bra{s,s}\hat{V}_{dd}\ket{s,s} = 0$, but can interact via second-order vdW interactions $V_{\text{vdW}} = C_6/r^6$. 

\begin{figure}[tb!]
\centering
\includegraphics[width=\columnwidth]{"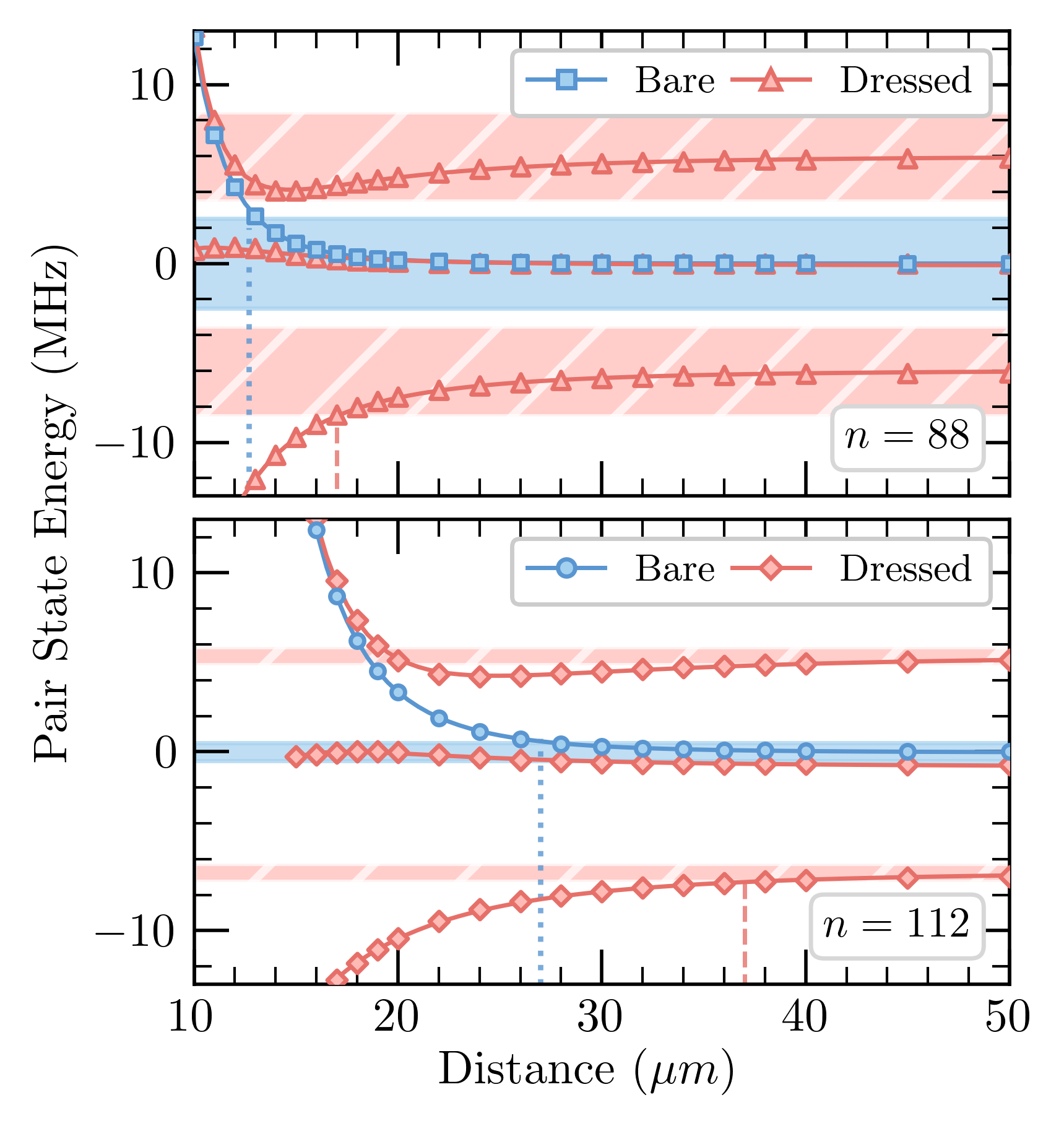"}
\caption{\label{fig:Fig3} Calculated interaction potentials for bare (blue) and dressed (red) two-atom states oriented at $0^\circ$ with respect to the quantization axis. The lines through the points are fits to $C_6/r^6 + C_3/r^3$. The bands around the large-$r$ energies of the $\ket{ss}$  and $\ket{--}$ states correspond to $\pm \Omega_\text{Ry}$. The blue dotted (red dashed) vertical lines indicate the blockade radii for the bare (dressed) states.}
\end{figure}

By admixing opposite-parity Rydberg levels, stronger interactions can be generated.
To first order, the strength of the interactions for the dressed two-atom states,  $\ket{++}$ and $\ket{--}$, can be written as 
\begin{align}
   & \bra{--} \hat{V}_\text{dd} \ket{--}  = 
    \bra{++} \hat{V}_\text{dd} \ket{++} \nonumber \\  &\qquad = \frac{1}{4}\left[\bra{sp}\hat{V}_\text{dd} \ket{ps} + \bra{ss}\hat{V}_\text{dd} \ket{pp} + \text{H.c.} \right].
\end{align}
First-order pair interactions are anisotropic, switching from maximally attractive for pairs aligned with the quantization axis to maximally repulsive for pairs transverse to the axis of the cloud.
The combined DD and vdW potential for the dressed states therefore has regions transverse to the cloud where the dressing weakens the interactions compared to the bare vdW potential. 
However, the transverse size of the Rydberg ensemble in our experiment is defined by the probe, which is much narrower than both the longitudinal waist of the cloud and the vdW blockade radii $r_b$ and very few of the atom pairs in our cloud sample the region of weakened interactions, making the enhanced blockade effective for our quasi-1D ensemble.

In our experiment, the microwave drive and the DD interactions near $r_b$ are of comparable energy and both induce couplings between states in the bare basis.
The dressed states cannot be computed perturbatively so  we use the Floquet formalism to calculate the energies \cite{Shirley.1963}. 
The results are shown in Fig. \ref{fig:Fig3}, where the energy shifts of the bare and dressed two-atom states are plotted as a function of the pair separation distance.
The  states $\ket{++}$ and $\ket{--}$ have stronger interactions than $\ket{ss}$, resulting in a larger blockade radius (defined as the distance $r_b$ for which the collectively enhanced Rydberg excitation Rabi frequency equals the interaction strength, i.e.~$\left|V(r_b)\right| = \Omega_\text{Ry} \equiv \sqrt{N(r_b)}\Omega_p \Omega_c / 2 \Delta$.
The distance where the interaction shifts exceed $\pm\Omega_\text{Ry}$ (computed for our highest-density clouds with the greatest enhancement of the Rabi frequency) is shown as distance at which the potentials venture outside the blue and red bands.
Notice that the blockade radii for $n=88$ are much lower than $n=112$, in part due to the significant difference between the $\Omega_\text{Ry}$ used.
Due to the $1/r^3$ nature of the DD interactions, the enhancement of interactions and blockade radii are most evident at large distances and small $\Omega_\text{Ry}$. For example, when $\Omega_\text{Ry}/(2 \pi) < 0.7$ MHz, the blockade radius for the $n=88$ dressed state is larger than that of $n=112$ bare state despite the significant difference in $n$~\cite{Supplemental}.

{\it Theoretical model }---
Modeling Rydberg ensemble systems in the intermediate-blockade regime is a challenging task \cite{Berman.2024}. 
We adopt a 1D ``pseudo-atom model’’ \cite{Sun.2008}, where the ensemble is partitioned along the longitudinal direction into small bins such that pairs of atoms in one bin are perfectly blockaded.
The collection of atoms in each bin are then treated as a single pseudo-atom, modeled as a two-level pseudo-spin driven by $\Omega_\text{Ry}$, where the enhancement is determined by the number of atoms in the bin.
To model the atom cloud, we perform calculations for chains up to 60 pseudo-spins 
and truncate the Hilbert space by projecting our dynamics onto the set of states with up to 3 simultaneous excitations. 
The interactions between pseudo-spins are modeled according to the Floquet potentials presented in Fig.~\ref{fig:Fig3}.
To account for interaction-induced dephasing during the write and hold stages, we calculate the collective spin-wave pair-correlation function \cite{Bariani.2012v2} via Monte Carlo simulations where the atoms are sampled from the measured density profiles \cite{Supplemental}.
The results of our model with no free parameters are shown in Fig.~\ref{fig:Fig2}(c,d), demonstrating good agreement with the experimental results across all datasets. For a discussion of the small deviation of the simulations from the experimental results, see SM~\cite{Supplemental}.

Because $g^{(2)}(0)$ is an indirect measure of the interactions, there are alternative mechanisms that could explain its reduction. One is a possible $\Omega_\text{Ry}$-dependence of $g^{(2)}(0)$, since the dressed-state admixture of $nP_{3/2}$ decreases $\Omega_\text{Ry}$ from the bare value. We show experimentally~\cite{Supplemental} that the microwave-induced change in $\Omega_\text{Ry}$ has a negligible effect on the suppression of $g^{(2)}(0)$.

Another mechanism is the microwave enhancement of interaction-induced dephasing during storage \cite{Bariani.2011,Bariani.2012}. 
Previous experiments have explored this effect during extended hold times, observing that multi-excitation components dephase more rapidly with the application of microwaves, reducing the $g^{(2)}(0)$ of the retrieved light \cite{Maxwell.2013,Maxwell.2014,Xu.2024}.
To study the impact of dephasing during the hold period, we simulate the $g^{(2)}(0)$ of the retrieved light when the excitations are retrieved immediately after the write sequence. We find that multi excitations dephase over time scales $\sim\Omega_\text{Ry}^{-1}$. Consequently, the short storage time has a negligible impact on $g^{(2)}(0)$ for the $n=112$ results, but dephasing has a non-zero effect for $n=88$ since $\Omega_\text{Ry}^{-1} \approx t_s$. However, even for the simulations with no storage time, microwave dressing substantially reduces $g^{(2)}(0)$ for both principal quantum numbers, indicating that dephasing during the storage period alone cannot account for the enhancement of the single-photon purity. For a detailed discussion of both effects, see SM~\cite{Supplemental}. 

Our results indicate that direct blockade is the predominant mechanism for our reduced $g^{(2)}(0)$. In contrast to suppressing $g^{(2)}(0)$ through dephasing alone, which is a filtering process with maximum efficiency $1/e$ \cite{Bariani.2011}, blockade allows for the generation of high-purity single photons without fundamental limits to the source efficiency  \cite{Ornelas-Huerta.2020,Shi.2022,Yang.2022}. 

{\it Conclusion }---
In summary, we observe a significantly reduced $g^{(2)}(0)$ for light generated using microwave-dressed Rydberg states due to the enhancement of the blockade radius. 
Using  pure microwave polarizations we addressed specific Zeeman states and performed high fidelity Rydberg dressing. 
We measure $g^{(2)}(0)$ with and without dressing with varying cloud sizes to characterize the resulting  interaction strengths, and find good agreement with our model using Floquet interaction potentials and Monte Carlo simulation of the density profile effects. 

This work enables more ambitious interaction-engineering proposals.
Our platform can utilize different microwave polarizations and detunings as well as multiple microwave frequencies, which may allow the generation of long-range bound molecules and dominant three-body interactions \cite{Sevincli.2014}.  
Other proposals have shown that choosing appropriate detunings and polarizations can generate microwave-dressed eigenstates with nonvanishing interspecies and vanishing intraspecies interactions \cite{Young.2021} or create dark states with nullified vdW interactions \cite{Shi.2017}.
Expanding control over microwave fields to tune the properties of Rydberg atoms will permit the realization of novel applications in quantum optics, communication, and computation.  

\textit{Acknowledgments }---We thank Ian Spielman, and Alicia Kollar for their critical review of the manuscript. D.K., P.R.B., and Y.L. were supported by the MAQP grant W911NF2420107. S.R.M. is supported by the NSF QLCI grant OMA-2120757. A.V.G.~was supported in part by DARPA SAVaNT ADVENT, AFOSR MURI, DoE ASCR Quantum Testbed Pathfinder program (awards No.~DE-SC0019040 and No.~DE-SC0024220), NSF QLCI (award No.~OMA-2120757), DoE ASCR Accelerated Research in Quantum Computing program (awards No.~DE-SC0020312 and No.~DE-SC0025341), and the NSF STAQ program. A.V.G.~also acknowledges support from the U.S.~Department of Energy, Office of Science, National Quantum Information Science Research Centers, Quantum Systems Accelerator.  

\widetext
\clearpage

\begin{center}
\textbf{\large Supplemental material for Enhancement of Rydberg Blockade via Microwave Dressing}
\end{center}

\begin{center}
Deniz Kurdak\,\orcidlink{0000-0003-4076-3013},$^{1}$ Patrick R. Banner\,\orcidlink{0009-0006-9957-4996},$^{1}$ Yaxin Li\,\orcidlink{0000-0001-8734-0136},$^{1}$ Sean~R.~Muleady\,\orcidlink{0000-0002-5005-3763},$^{1,2}$ \\[3pt] Alexey V. Gorshkov\,\orcidlink{0000-0003-0509-3421},$^{1,2}$ S. L. Rolston\,\orcidlink{0000-0003-1671-4190},$^{1}$ ,J. V. Porto\,\orcidlink{0000-0002-6290-7535}$^{1}$
\end{center}

\begin{center}
\textit{$^1$ \JQIalt} \\
\textit{$^2$\QUICSalt}
\end{center}

\newenvironment{myquote}%
  {\list{}{\leftmargin=0.6in\rightmargin=0.6in}\item[]}%
  {\endlist}

\begin{myquote}
\hspace{6pt} In this supplemental material, we highlight additional details of our work. In Sec. \ref{sec:cloudShaping}, we discuss the experimental techniques utilized to vary the length of our ensemble. In Sec. \ref{sec:MicrowaveTuning}, we elaborate on the details of our microwave purification techniques, including the spectroscopy methods and microwave equipment used. In Sec. \ref{sec:Floquet}, we present further results of the pair-interaction calculations, illustrating the dependence of the dressed-interaction potentials on the angle between the interatomic and quantization axis. We also calculate the enhancement of the blockade radii via dressing for a variety of Rydberg excitation Rabi frequencies. In Sec. \ref{sec:theory_model}, we elaborate on the theoretical model of the Rydberg excitation and dephasing dynamics in our ensemble. We define the observables used to capture $g^{(2)}(0)$ and the approximations made to simplify the computational complexity of the simulations. And lastly, in Sec. \ref{sec:Dephasing}, we discuss the impact of the control Rabi frequency and storage period on the statistics of the retrieved light.
\end{myquote}
\setcounter{equation}{0}
\setcounter{figure}{0}
\setcounter{table}{0}
\setcounter{page}{1}
\makeatletter
\renewcommand{\theequation}{S\arabic{equation}}
\renewcommand{\thefigure}{S\arabic{figure}}
\renewcommand{\bibnumfmt}[1]{[S#1]}
\renewcommand{\citenumfont}[1]{S#1}

\section{Cloud Shaping}\label{sec:cloudShaping}

\begin{figure}[t!]
    \centering
    \includegraphics{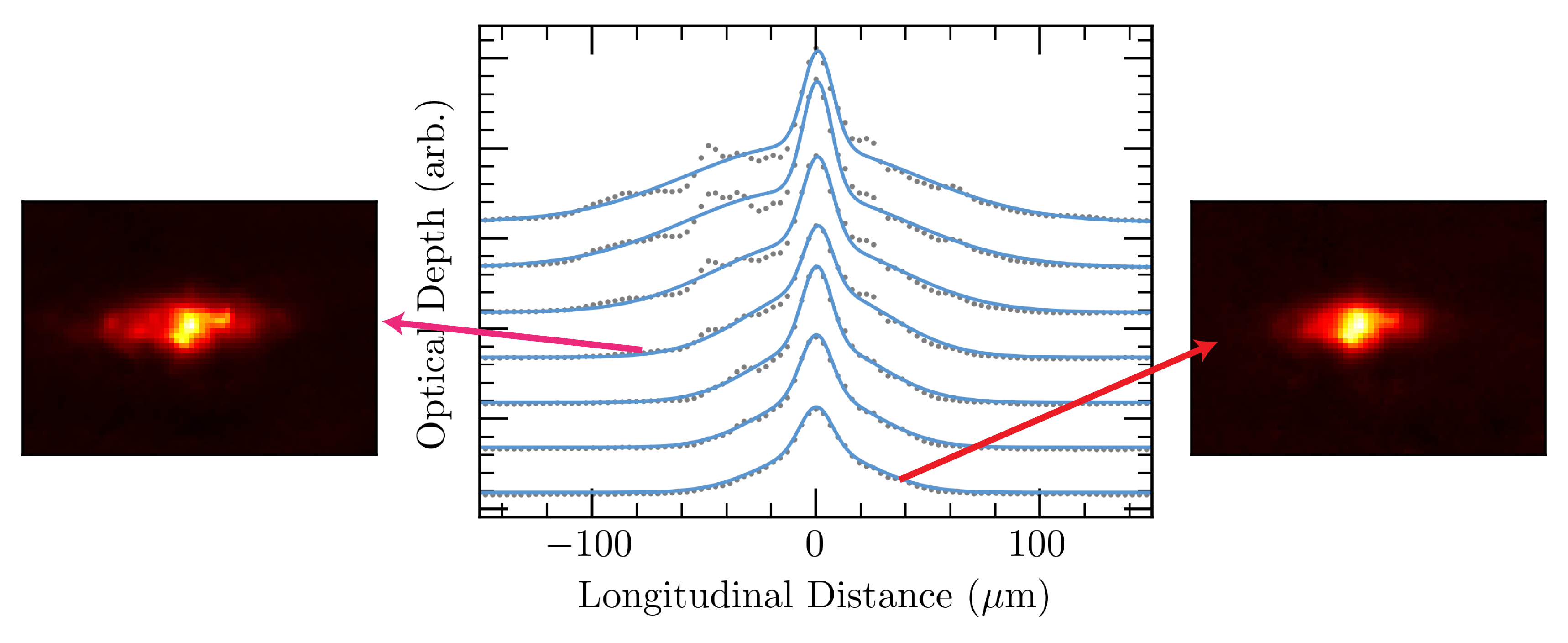}
    \caption{ Tunability of the cloud waist along the probe direction. The left and right panels show absorption images of clouds with different longitudinal RMS radii. The images are taken perpendicular to the probe beam, which travels from left to right along the long direction of the cloud. The center panel shows the longitudinal cross-sections extracted from absorption images of clouds of different lengths. The cross-sections have been vertically offset for visual clarity. The cloud length is adjusted by varying the depth of the crossed trap during the cloud shaping procedure. The longitudinal profiles are fit to a double Gaussian, \ref{eq:eqS1}. The peak density of the ensemble increases with cloud length, where the shortest and longest clouds have a density of $\approx 2 \times 10^{11}$ cm$^{-3}$ and $\approx 5 \times 10^{11}$ cm$^{-3}$, respectively.
    \label{fig:SupFig1}}
\end{figure}

In this section, we discuss the methods used to confine the atoms in a 1D lattice with a variable longitudinal root-mean-square (RMS) radius, as shown in Fig.~\ref{fig:SupFig1}. The atoms are transferred from the MOT into a dipole trap composed of a crossed pair of beams and a transverse beam. The crossed trap produces a cigar-shaped cloud with a longitudinal RMS radius $\approx 50 \, \mu$m. Conversely, the transverse beam is tightly focused to a waist of nearly $\sim 20 \, \mu$m. The polarization of the retroflected arm of the crossed trap is controlled by a liquid-crystal variable retarder (LCR). During dipole-trap loading, the polarization of the in-going and retro arms of the crossed trap are orthogonal. After the loading and cooling stages, the crossed trap is lowered adiabatically to a variable depth over 10 ms while transverse beam is kept at full power. The atoms are held for 30 ms, where they either leave the trap or fall in the potential modified by the small transverse beam. The crossed trap depth used during this stage determines the length of the final cloud, where keeping the crossed trap fully on realizes the longest cloud configuration and turning it off fully realizes the minimum cloud length determined by the waist of the transverse beam. The atoms are held for an additional 30 ms as the LCR rotates the polarization of the retro beam to create a 1D lattice. The crossed trap depth is slowly ramped back up to its final value over 10 ms, and the transverse beam is turned off to avoid vector light shifts during the Rydberg excitation sequence. The crossed trap provides the transverse confinement, and the 1D lattice restricts the relaxation of the cloud in the longitudinal direction, fixing the cloud length. 

In Fig.~\ref{fig:SupFig1}, we show two representative absorption images of our cloud as well as the longitudinal cross section of a variety of clouds accessed by using a variable crossed trap depth during the cloud shaping procedure. The fits in the center plot are double Gaussians, 
\begin{equation}
    \text{OD}(x) = a_1 \exp{-(x-o_1)^2/2w_1^2} + a_2 \exp{-(x-o_2)^2/2w_2^2} + c \label{eq:eqS1}
\end{equation}
From the fits, we extract the amplitude and waist of the two Gaussians to estimate the fraction of atoms hosted by the narrow and broad distributions. To represent the cloud shape by a single parameter we define an effective cloud waist, which is given by the weighted quadrature sum over the two Gaussians: 
\begin{equation}
    w_\text{eff} = \sqrt{\frac{a_1 w_1^3 + a_2 w_2^3}{a_1 w_1 + a_2 w_2}}.
\end{equation}
We find that the effective waist tracks the waist of the larger Gaussian closely, meaning the narrow Gaussian hosts a small fraction of the total number of atoms. Furthermore, the narrow Gaussian has a small enough waist that the blockade radii considered in this work cover it effectively. Therefore, the bimodal distribution of the optical depth has no practical impact on the results discussed in the main text. In our experiment, the mismatch of the centers of the two Gaussian distributions is small, $o_1 - o_2 \ll w_1,\, w_2$, meaning $w_\text{eff}$ is equivalent to the RMS radius of the ensemble.

\section{Tuning of the Microwave Polarization}\label{sec:MicrowaveTuning}
\begin{figure}[b!]
    \centering
    \includegraphics{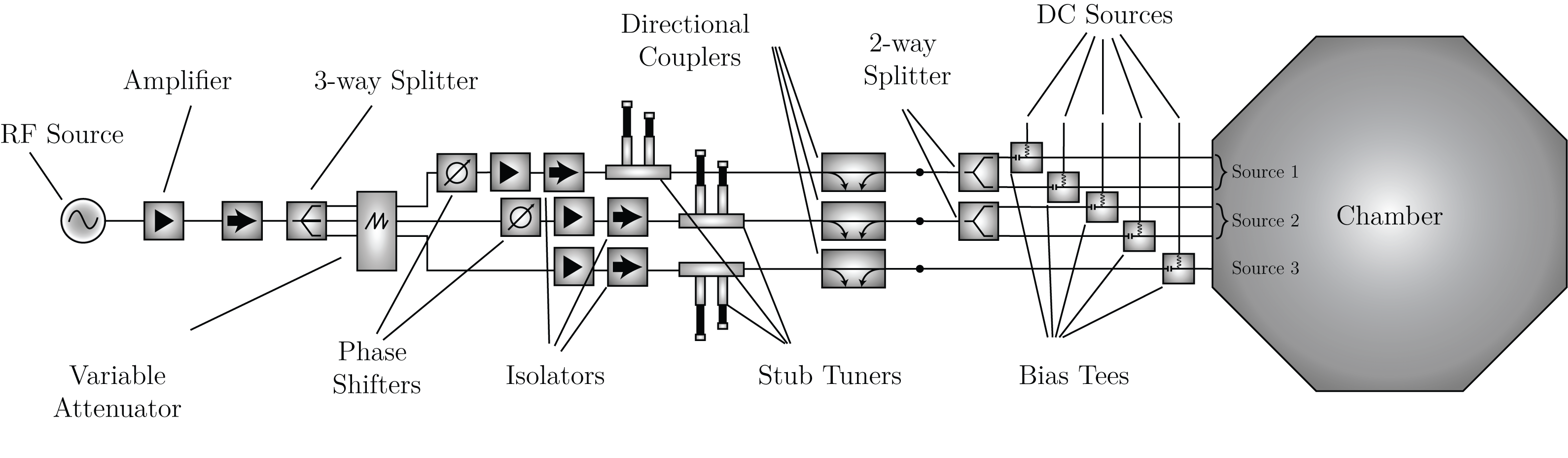}
    \caption{The microwave equipment used to drive three independent sources. A single source is split into three channels with arbitrary amplitude and relative phase control. Stub tuners are used to achieve a constant response over the frequency range of interest. The microwaves are coupled to the in-vacuum electrodes that also serve as the DC bias field controls; the microwave and DC fields are combined with bias tees before entering the chamber}
    \label{fig:SupFig2}
\end{figure}

In this section, we describe the microwave equipment and spectroscopy techniques used to tune the polarization of the microwave fields at the position of the atoms.
As shown in Fig.~1(a) of the main text, we use three sets of electrodes as linearly independent sources to adjust the polarization of the microwave fields. Note that these electrodes are intended for DC electric field control and are adapted to provide microwave fields.
The microwave system used to drive the three independent sources is shown in Fig.~\ref{fig:SupFig2}.
The microwaves, generated by a single RF source, pass through an amplifier and isolator before being split into three independent sources. 
The RF power of each source is controlled by a variable attenuator and the relative phases between the three sources are controlled using two variable phase shifters on Sources 1 and 2.
As outlined in the main text, we utilize in-vacuum electrodes both for canceling out stray electric fields and for applying microwave fields. 
As such, the DC and AC sources are combined with a set of bias tees before being sent into the chamber via a D-Sub 9 feed-through. 
To minimize the microwave reflections from the chamber and to ensure constant spectral response over the frequency range of operation, each source was impedance matched using stub tuners. 

\begin{figure}[t!]
    \centering
    \includegraphics{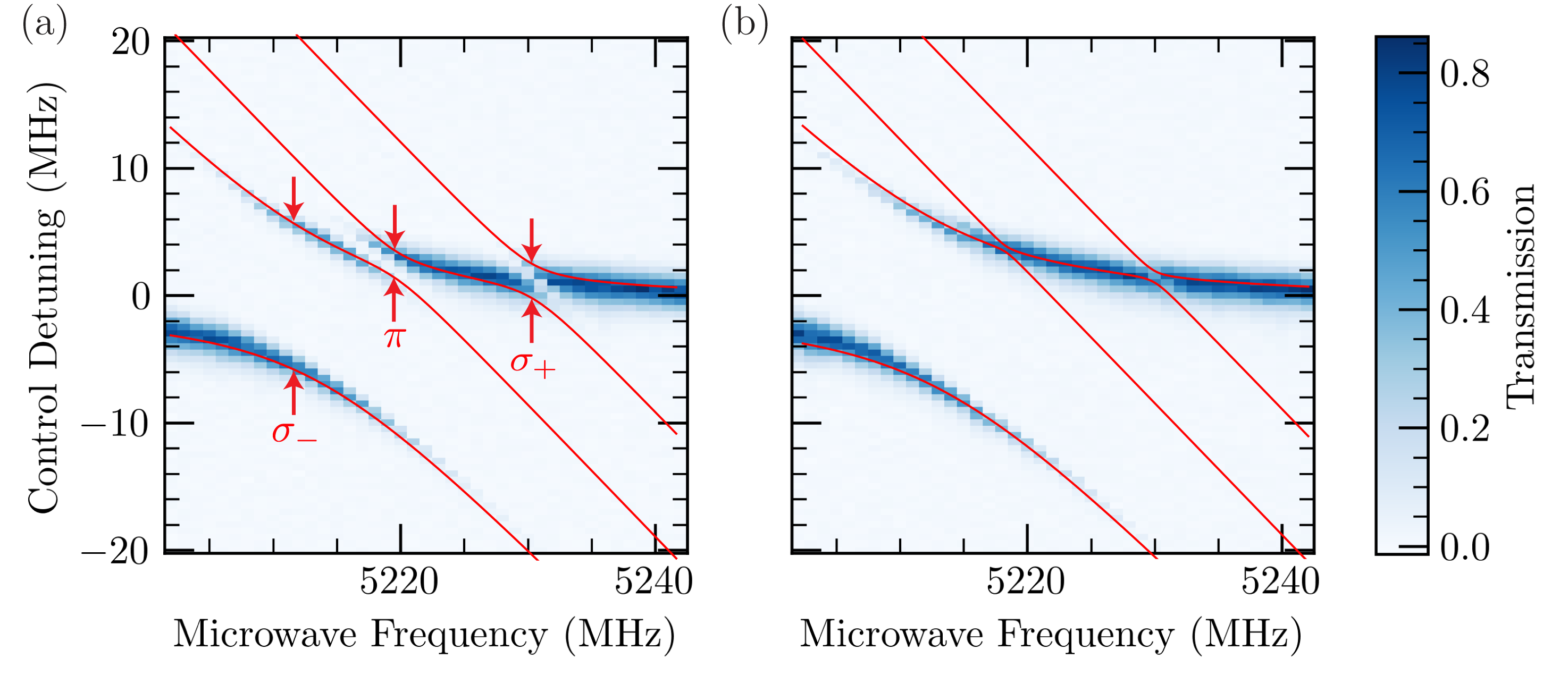}
    \caption{EIT spectroscopy of the microwave induced avoided crossings for $88S_{1/2}$. The probe transmission on resonance was monitored as the control and microwave frequencies were swept. Figures (a) and (b) correspond to the spectra taken before and after the microwave polarizations were purified to the $\sigma_-$ state. The red arrows, labeled by the relevant microwave polarizations, indicate the positions of the three avoided crossings. The red lines indicate fits using a four-level model from which the microwave Rabi frequencies were extracted.} 
    \label{fig:SupFig3}
\end{figure}

A magnetic field of 0.5 mT was applied to split the $m_J$ levels of the $nP_{3/2}$ manifold, breaking the degeneracy of the transitions driven by the three polarizations. 
The probe and control lasers were used to perform electromagnetically induced transparency (EIT) spectroscopy: with the probe on resonance with the intermediate state, the control detuning was swept to locate the transmission windows indicating the presence of an eigensatate with a nonzero admixture of the $nS_{1/2}$ state.
The EIT sweeps were performed over a range of microwave frequencies \cite{Robinson.2021} covering the three resonances corresponding to the $\sigma_-$, $\pi$, and $\sigma_+$ microwaves.
On resonance, the microwave fields split the $nS_{1/2}$ levels by the microwave Rabi frequency. 
Therefore, the resonance frequencies as well as the microwave Rabi frequencies can be measured by mapping out the avoided crossings that occur in the control and microwave frequency space.
Fig.~\ref{fig:SupFig3}(a) shows the three avoided crossings observed for a microwave with all three polarization states.
We adjusted the microwave polarization by independently controlling the phase and amplitude of the three  sources, achieving a nearly perfect $\sigma_-$ polarized microwave drive. 
We used the $\sigma_-$ polarized microwaves in our experiment since the corresponding transition had the largest Clebsch-Gordan coefficient of the three available dressing transitions. 
Fig.~\ref{fig:SupFig3}(b), we show the avoided crossing measurement after the polarization tuning showing the near complete closing of the $\pi$ and $\sigma_+$ avoided crossings.

The red lines in Fig.~\ref{fig:SupFig3} correspond to a fit to the centers of the transmission windows extracted from the full transmission data. The fit model is a minimal $4 \times 4$ Hamiltonian that accounts for the Zeeman shifts, the microwave drives and the $s-p$ transition energy. The microwave Rabi frequencies and the unperturbed $s-p$ resonance frequency are free fit parameters. After tuning, at $n=88$, the microwave Rabi frequencies were $\Omega_{\sigma_-}/(2\pi), \, \Omega_\pi/(2\pi), \, \Omega_{\sigma_+}/(2\pi) = 12.5(2), \, 0.1(1), \, 0.8(2) \,$ MHz, where the numbers in the parentheses indicate the fit uncertainties. We can define a microwave tuning fidelity metric through 
\begin{equation}
    \mathcal{F} = \frac{E_{\sigma_-}}{\left|\mathbf{E}\right|} = \frac{\Omega_{\sigma_-}}{\sqrt{\Omega_{\sigma_-}^2 + \left(\Omega_{\pi}/\sqrt{2/3}\right)^2 + \left(\Omega_{\sigma_+}/\sqrt{1/3}\right)^2}} =99.3(3)\%.
\end{equation}
The scaling factors in the denominator are the Clebsch-Gordan coefficients that relate the Rabi frequencies to the electric field amplitudes. The error bound correspond to the 68\% confidence interval. In addition to the excellent field purity, the presence of the Zeeman detunings ensures that the undesired $m_J$ states have negligible contributions to the dressed state created.

\section{Pair Interaction Strength Calculations Using the Floquet Formalism}\label{sec:Floquet}
\begin{figure}[b!]
    \centering
    \includegraphics{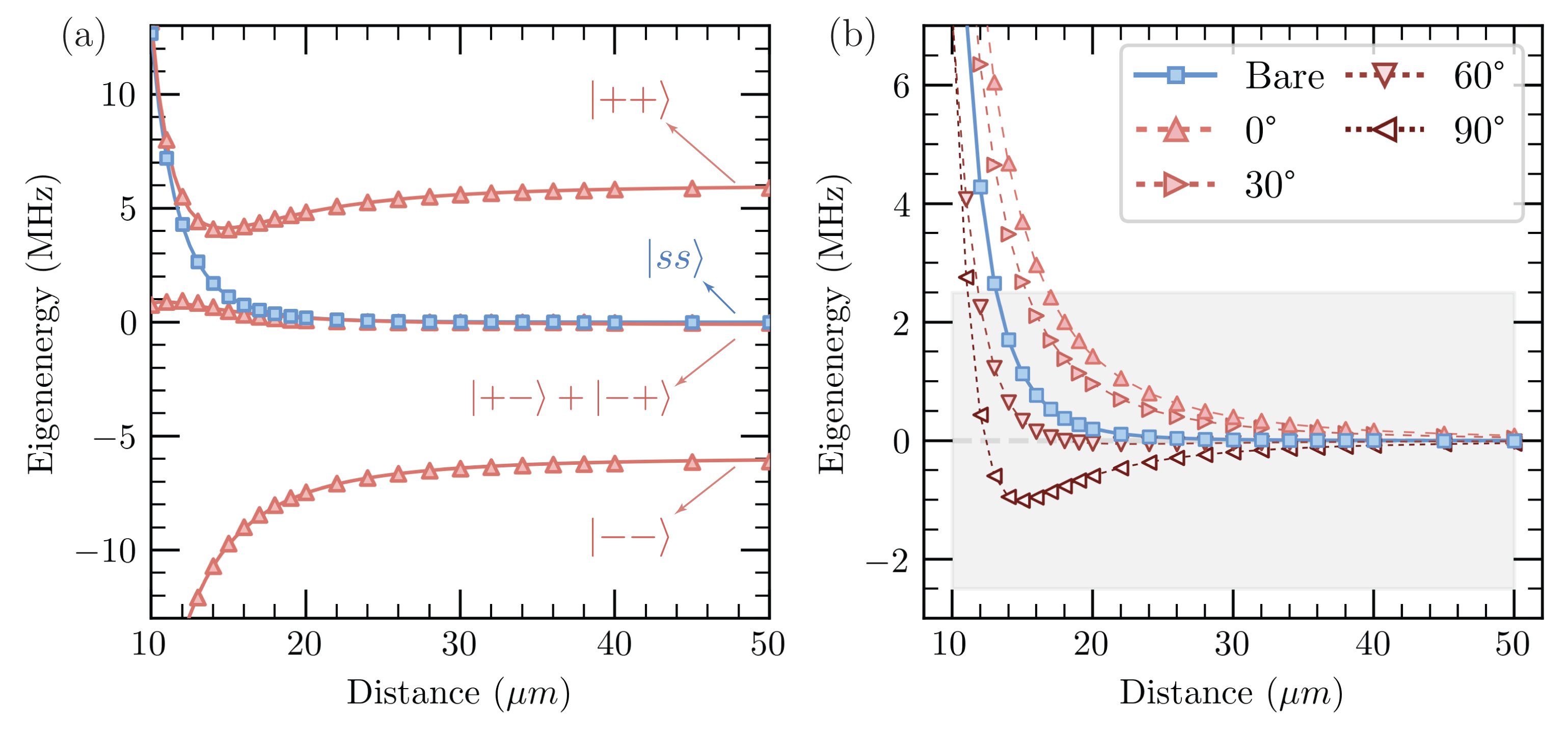}
    \caption{Floquet calculation of the pair-state eigenenergies as a function of pair separation. (a) The energies of the $n=88$ bare and dressed pair states at zero-degree orientation. The blue squares indicate the van der Waals potential of the bare state. The red triangles show the level shifts of the dressed pair states. The lower energy $\ket{--}$, which shows the largest enhancement of the interaction strength, was used for the results presented in the main text. The lines going through the plots are fits of the form $C_3/r^{3} + C_6/r^6$. (b) The anisotropy of the interactions for the $\ket{--}$ state. To facilitate easier comparison, the sign of the dressed-state energies has been flipped and asymptotic energy of the eigenstates has been subtracted. The blue squares and red triangles indicate the bare and dressed pair potentials, as indicated by the legend.}
    \label{fig:SupFig4} 
\end{figure}

The microwave Rabi frequencies used in the experiment and the interaction strength near the blockade radius were comparable in scale, making it difficult to treat either effect perturbatively. 
Therefore, the Floquet formalism was used to diagonalize the two-atom Hamiltonian \cite{Shirley.1963}. The truncated Hilbert space included pair states with $n\in\left[n-2,n+1\right]$, $l=0,1$, and $\abs{q}=0,1,2$, totaling 5120 two-atom Floquet states, sufficient for convergence of the calculations.

The results of the calculations at $n=88$ are presented in Fig.~\ref{fig:SupFig4}. 
Figure \ref{fig:SupFig4}(a) shows the impact of resonant ($\Omega_\mu/(2 \pi) = 6$~MHz) microwave dressing on the interaction of a pair of Rydberg atoms. 
The van der Waals potential of the pair of atoms in the $\ket{ss}$ state is indicated by the blue squares. Microwave dressing generates three eigenstates with non-negligible $\ket{ss}$ fractions, indicated by the red triangles. 
For large pair separations, the upper (lower) energy curve corresponds to the $\ket{++}$ ($\ket{--}$) state, as defined in the main text, while the center curve corresponds to the $\ket{ss} - \ket{pp} $ state. 
Note that, near $15 \, \mu$m pair separation, an avoided crossing between the upper and central dressed states opens due to the repulsive second-order van der Waals interactions.
The resulting deformation of the $\ket{++}$ state leads to the degradation of the enhanced blockade. 
Consistent with this prediction, we experimentally observe that the use of the $\ket{--}$ state leads to a greater suppression of $g^{(2)}(0)$ (data not shown). 

As suggested in the main text, the dressing of the bare Rydberg states breaks the isotropy of the van der Waals interactions between a pair of atom in the $\ket{s}$ state. 
In Fig.~\ref{fig:SupFig4}(b), the interaction strength of the $\ket{--}$ state is plotted as a function of the angle between the interatomic axis and the quantization axis set by the magnetic field. 
To better compare with the bare-state energy curve, the asymptotic energies of the pair states have been subtracted, and the dressed state energy curves have been flipped about the x-axis.
The gray band corresponds to the largest $\pm\Omega_\text{Ry}$ used in our experiment, such that the distance at which the interaction potentials venture outside of the band indicates the blockade radius. 
The enhancement of the interaction strength between a pair of atoms in the $\ket{--}$ state is strongest along the zero degree angle, and monotonically decreases as the angle is increased to $90^\circ$.
Notably, near the zero interaction angle of the pure dipole-dipole interactions ($\approx 54.7^\circ$)  \cite{Ravets.2014}, the dressed state energy curve approaches that of the bare states. 
Past this angle, the interaction strength is slightly reduced but remains nonzero.
As discussed in the main text, we utilize a cloud geometry where the interatomic axis of most pairs of atoms is aligned with the quantization axis such that the anisotropy of the dressed state interactions does not significantly affect our experiment.
Across all of our datasets, the ratio of pairs added to the blockaded region over those that were removed ranges between $\sim 6 - 30$. 

\begin{figure}
    \centering
    \includegraphics{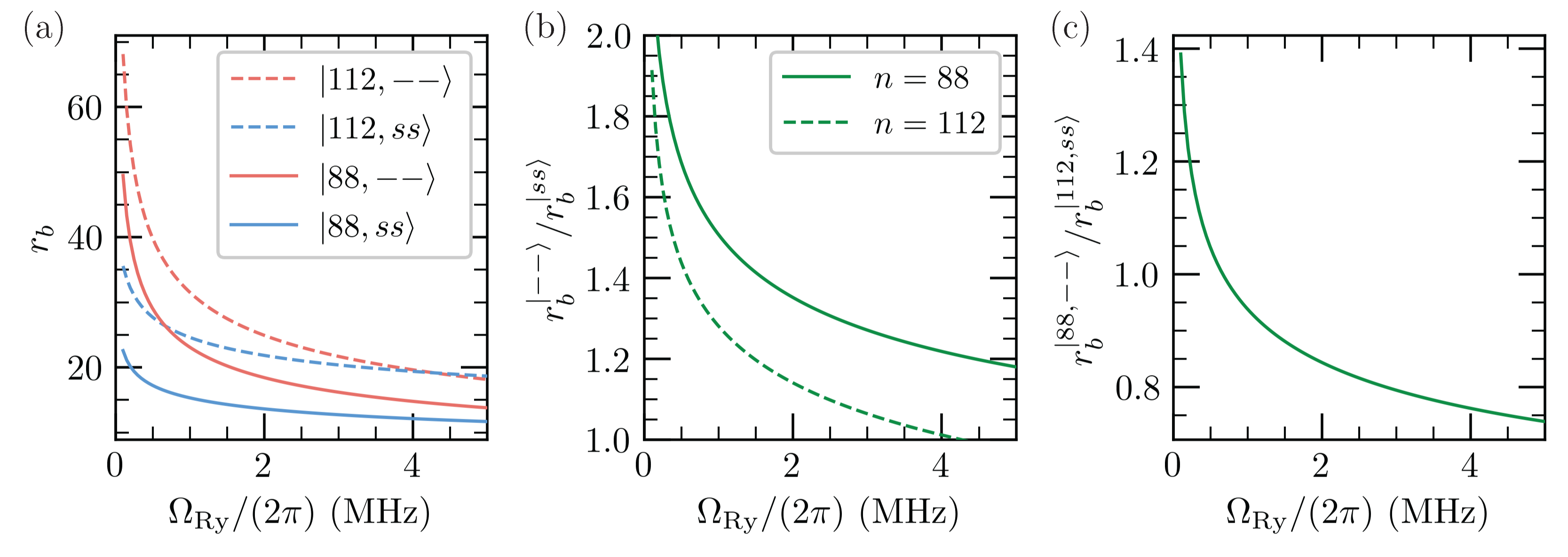}
    \caption{The blockade radii $r_b$, as a function of the collectively-enhanced Rydberg excitation Rabi frequencies, $\Omega_\text{Ry}$. The blockade radii are computed using the zero-angle Floquet potentials in Fig.~\ref{fig:SupFig4}, and self-consistently through $V(r_b) = \sqrt{N(r_b)} \Omega_s \equiv \Omega_\text{Ry}(r_b)$. (a) The blockade radii for the bare and dressed states at the two principal quantum numbers studied in this work. The dressed states are plotted in red, while the bare states are plotted in blue. The dashed lines and solid lines indicate the $n=112$ and $n=88$ blockade radii, respectively. (b) The ratio of the dressed and bare blockade radii at $n=112$ and $n=88$ for a range of $\Omega_\text{Ry}$. (c) The comparison of the bare $n=112$ blockade radii with the dressed $n=88$ blockade radii.}
    \label{fig:SupFig5}
\end{figure}

To further illustrate the enhancement of the interaction strength, we study the blockade radii achievable via dressing as a function of the Rydberg excitation Rabi frequency, as shown in Fig.~\ref{fig:SupFig5}. Fig.~\ref{fig:SupFig5}(a) shows the variation of the blockade radii with the Rydberg excitation Rabi frequency. Fig.~\ref{fig:SupFig5}(b) illustrates the significant enhancement of the blockade radius under dressing. At $2 \pi \times 1$ MHz Rydberg excitation Rabi frequency, the blockade radius is enhanced by $\approx 25 \%$ and $\approx 50\%$ for $n=112$ and $n=88$, respectively. Due to the very strong scaling of the van der Waals interaction strength with $n$, the enhancement of the blockade radius due to dressing is expected to provide increasing gains with decreasing principal quantum numbers. In Fig.~\ref{fig:SupFig5}(c), we compare the blockade radii achievable with dressing at $n=88$ and without dressing at $n=112$. Despite the significant difference in principal quantum numbers, there is a broad range of $\Omega_\text{Ry}$ over which the blockade radii are comparable. This observation is also consistent with the data presented in Fig. 2(b) of the main text, further highlighting that the results presented are in the regime dominated by the blockade physics.

\section{Theoretical Model}\label{sec:theory_model}
In this section, we model a cloud of $^{87}$Rb atoms held in a crossed dipole trap driven to a Rydberg level through a two-photon transition. Such levels may be dressed through the use of polarization-controlled microwave beams, leading to modifications to the bare Rydberg atom potential. As depicted in Fig~1(c) of the main text, for each atom, we have a ground state $|g\rangle$ and an intermediate excited state $|e\rangle$ which is optically coupled to the ground state via an off-resonant probe beam, with Rabi frequency $\Omega_{\mathrm{p}}(\mathbf{r})$ and detuning $\Delta$. $|e\rangle$ is also coupled off-resonantly to a Rydberg state $|\mu\rangle$ by a control beam with Rabi frequency $\Omega_{\mathrm{c}}(\mathbf{r})$ and detuning $-\Delta$. This realizes an effective resonant two-photon drive between $|g\rangle$ and the Rydberg state $|\mu\rangle$.

In the absence of microwave driving, the relevant Rydberg state is an $s$ state $|\mu\rangle = |s\rangle$. The relevant single-atom Rabi frequency for the effective two-photon drive is given via $\Omega_{\mathrm{eff},s}(\mathbf{r}) = \Omega_{\mathrm{c}} (\mathbf{r})\Omega_{\mathrm{p}}(\mathbf{r})/(2\Delta)$. In the case that this $s$ state is strongly coupled to a $p$ state by a resonant microwave drive, the Rydberg level corresponds to an effective dressed state $|\mu\rangle = (|s\rangle \pm |p\rangle)/\sqrt{2}$. As the optical two-photon drive only couples to the $s$ state, the effective single-atom Rabi frequency in this case is $\Omega_{\mathrm{eff},\pm}(\mathbf{r}) = \Omega_{\mathrm{c}}(\mathbf{r}) \Omega_{\mathrm{p}}(\mathbf{r})/(2\sqrt{2}\Delta)$. We thus model the two-photon drive via the following Hamiltonian:
\begin{align}
    \hat{H}_{\mathrm{drive},\mu} = \sum_i \frac{\hbar\Omega_{\mathrm{eff},\mu}(\mathbf{r}_i)}{2}\left(|\mu\rangle_{i}\langle g|_i + |g\rangle_{i}\langle \mu|_i\right),
\end{align}
where the label $\mu$ denotes the relevant Rydberg state, and the index $i$ runs over all atoms in the cloud, which have corresponding positions $\mathbf{r}_i$. We note that we include the spatial dependence of the beam envelope over the atom cloud in $\Omega_{\mathrm{eff},\mu}(\mathbf{r}_i)$. However, we assume the spatial phase for each atom, $e^{i\mathbf{k}\cdot\mathbf{r_i}}$ where $\mathbf{k}$ is the wavevector for the two-photon drive, is absorbed into the definitions of $|\mu\rangle_i$, $|g\rangle_i$, so that $\Omega_{\mathrm{eff},\mu}$ is real.

Pairs of atoms in the $|s\rangle$ and $|\pm\rangle$ states interact via the two-body interaction potentials $V_{ss}(r) = C_{6,ss}/r^6$ and $V_{\pm\pm}(r) = C_{6,\pm\pm}/r^6 + C_{3,\pm\pm}/r^3$, respectively. The $C_3$ and $C_6$ coefficients for the pair interaction potentials are acquired from fits to the Floquet simulations presented in Sec. \ref{sec:Floquet}. The above potentials well approximate the interaction strength for distances outside the so-called spaghetti region. 
For very short distances,  pairs are well-blockaded and the exact form of the interaction used is irrelevant.  
Due to the nearly one-dimensional geometry of relevant atoms illuminated by the probe beam, we observe negligible angular variation in the interaction potentials between any relevant pair of atoms, and thus utilize these spherically symmetric potentials for simplicity. We can describe these interactions via the interaction Hamiltonian
\begin{align}
    \hat{H}_{\mathrm{int},\mu} = \sum_{i<j} \hbar V_{\mu\mu}(r_{ij}) |\mu\mu\rangle_{ij} \langle\mu\mu|_{ij},
\end{align}
where $r_{ij} = |\mathbf{r}_i - \mathbf{r}_j|$.

For modeling the experimental sequence and the resulting $g^{(2)}(0)$, we assume atoms to be initialized in the ground state, i.e.~we have initial state $|\psi_0\rangle = \bigotimes_{i} |g\rangle_i$. We apply our drive and interaction Hamiltonians for a write time $t_1$, resulting in $|\psi_1\rangle = \exp\{-it_1(\hat{H}_{\mathrm{drive},\mu} + \hat{H}_{\mathrm{int},\mu})/\hbar\}|\psi_0\rangle$. We then apply the interactions for an additional hold time $t_2$, during which the drive Hamiltonian is off, but the interactions continue to induce phases between particles, leading to $|\psi_2\rangle = \exp\{-it_2\hat{H}_{\mathrm{int},\mu}/\hbar\}|\psi_1\rangle$. To compute the pulse-integrated $g^{(2)}(0)$ following the retrieval from $|\psi_2\rangle$, we assume that the process of retrieval does not alter $g^{(2)}(0)$ compared to the $g^{(2)}$ of the excitations stored in the spin-wave mode. 
We thus have
\begin{align}\label{eq:g2}
    g^{(2)}(0) = \frac{\langle\psi_2| \hat{S}_0^+ \hat{S}_0^+ \hat{S}_0^- \hat{S}_0^-|\psi_2\rangle}{\langle \psi_2| \hat{S}_0^+ \hat{S}_0^-|\psi_2\rangle^2},
\end{align}
where $\hat{S}_0^\pm = \sum_i \hat{s}_i^\pm/\sqrt{\mathcal{N}}$  is the creation operator for the the collective spin-wave mode, with spin ladder operators $\hat{s}_i^+ = |\mu\rangle_i \langle g|_i$, $\hat{s}_i^- = (\hat{s}_i^+)^\dagger$ and normalization factor $\mathcal{N}$.

In computing these dynamics, we first form an ensemble of particle positions by Monte Carlo sampling from the experimentally characterized distribution. We also assume that the spatial variation of the write beam is determined by the relatively narrower probe beam, which is a Gaussian beam with waist $w_0 = 3.3\,\mu$m and Rayleigh range $z_R = \pi w_0^2/\lambda = 44 \,\mu$m for $\lambda = 780$ nm. If we assume the probe axis to be centered at $x=0$, $y=0$, then we have
\begin{align}
    \Omega_{\mathrm{eff},\mu}(\mathbf{r}) = \frac{\Omega_{\mathrm{eff},\mu}(\mathbf{0})}{\sqrt{1 + (z/z_R)^2}}\exp\left\{-(x^2 + y^2)/\left[w_0^2(1+z/z_R)^2\right]\right\}.
\end{align}
Besides the experimentally characterized trap geometry, we also utilize the same write and hold times, $t_1$ and $t_2$, used in the experimental sequence for our calculations. 
We numerically chose $\Omega_{\text{eff},\mu}$ such that the simulated excitation dynamics were consistent with the experimentally observed Rabi flops between the collective ground and excited states.

To realistically evaluate Eq.~\eqref{eq:g2}, we make two important simplifications. First, given that strong blockade effects effectively prevent more than a single atom within a blockade radius from being excited to $|\mu\rangle$, we project our dynamics onto the set of states with up to $k$ simultaneous excitations to the $|\mu\rangle$ state. The number of excitations we need to attain convergence of $g^{(2)}(0)$ in $k$ increases as the size of the ensemble increases. For the cloud lengths relevant to the experiment, we find that $k=3$ is sufficient to attain convergent behavior, by comparing with results using $k=4$. This ``few-polariton'' approximation vastly reduces the requisite Hilbert space dimension to $\sim N^k$, where $N$ is the number of atoms in the system.

The second approximation we make is to adopt a ``pseudo-atom model,'' which further expedites our calculations, though is not strictly necessary for the problem at hand (see Ref.~\cite{Sun.2008} for a similar approach). In this approximation, we sort particles into spatial bins, where the $J$-th bin contains all particles $i$ whose $z$ coordinate lies in the range $Jz_{\textrm{bin}} \leq z_i \le (J+1) z_{\textrm{bin}}$ for integer $J$, and where $z_{\textrm{bin}}$ sets the width of each bin. That is, if we denote the set of indices in the $J$-th bin via $\mathcal{B}_J$, then $\mathcal{B}_J = \left\{i|Jz_{\textrm{bin}} \leq z_i \le (J+1) z_{\textrm{bin}}\right\}$. If $z_{\textrm{bin}}$ is sufficiently small compared to the blockade radius, only one atom within each bin---or ``pseudo-atom''---can be excited to the Rydberg state at a time. Furthermore, such excitations will only exist within a symmetric superposition between all atoms in the bin. We can thus model all atoms in the $J$-th bin via a single spin-1/2 degree of freedom, where the relevant states are the ground state $|g\rangle_J \equiv |ggg...\rangle$ and the $W$ state $|\mu\rangle_J \propto |\mu ggg...\rangle + |g\mu gg... \rangle + |gg\mu g...\rangle + ...$. 

In reality, for a spatially varying beam profile, atoms will not be symmetrically excited to the $W$ state. To take this into account, we define an effective number of atoms $N_{\mathrm{eff},J}$ within the $J$-th bin, such that $N_{\mathrm{eff},J} = \sum_{i\in \mathcal{B}_J} (\Omega_{\mathrm{eff},\mu}(\mathbf{r}_i)/ \Omega_{\mathrm{eff},\mu}(\mathbf{0}))^2$, and we have total effective atom number $N_{\mathrm{eff}} = \sum_J N_{\mathrm{eff},J}$. To construct Hamiltonians that describe similar evolution of this pseudo-atom model, we first note that the Rabi frequency between the $|g\rangle_J$ and $|\mu\rangle_J$ atoms will exhibit a $\sqrt{N_{\mathrm{eff},J}}$ enhancement compared to the bare Rabi frequency. We thus have 
\begin{align}
    \hat{H}_{\mathrm{drive},\mu}^{\mathrm{PA}} = \sum_J \frac{\hbar\Omega_{\mathrm{eff},\mu}(\mathbf{0})\sqrt{N_{\mathrm{eff},J}}}{2}\left(|\mu\rangle_{J}\langle g|_J + |g\rangle_{J}\langle \mu|_J\right).
\end{align}
Next we define the pseudo-atom interaction Hamiltonian via
\begin{align}
    \hat{H}_{\mathrm{int},\mu}^{\mathrm{PA}} = \sum_{I<J} \hbar V_{\mu\mu,IJ}|\mu\mu\rangle_{IJ} \langle\mu\mu|_{IJ}.
\end{align}
Here, $V_{\mu\mu,IJ} = \sum_{i\in\mathcal{B}_I,j\in\mathcal{B}_J}V_{\mu\mu}(r_{ij})/N_{\textrm{eff},I}N_{\textrm{eff},J}$ is the average interaction over all unique pairs between bins $I$ and $J$. This average interaction between bins should remain valid as long as $z_{\textrm{bin}}$ remains sufficiently small. Lastly, for measuring spin wave observables of this pseudo-atom model, we have pseudo-atom matrix elements
\begin{align}
    \langle\mu|_J \hat{S}_0^+ |g\rangle_J = \sqrt{\frac{N_{\mathrm{eff},J}}{\mathcal{N}}}
\end{align}
for our pseudo-atom states. Thus, we can define our pseudo-atom spin wave observable via
\begin{align}
    \hat{S}_0^{+,\mathrm{PA}} = \sum_J\sqrt{\frac{N_{\mathrm{eff},J}}{\mathcal{N}}}|\mu\rangle_J \langle g|_J.
\end{align}

In practice, we find that our pseudo-atom model retains good convergence with the full (projected) Hamiltonian dynamics for $z_{\mathrm{bin}}$ as large as $10\,\mu$m; we utilize a value of $z_{\mathrm{bin}} = 5\,\mu$m for all our calculations. To model the full atom cloud, we include up to 60 pseudo-atoms in our model. 

As shown in Figure 2 of the main text, despite slightly overestimating the $g^{(2)}(0)$, the simulations are in good agreement with experimental results. The remaining minor discrepancy could be due to the absence of more subtle mechanisms not included in our model. One example is the dependence of the retrieval efficiency of the singly and doubly excited spin-waves on the spatial wavefunction of the spin-waves. While interaction effects during retrieval and the dependence of spin-wave spatial spectral properties on excitation number may have some effect on $g^{(2)}(0)$, our model assumes that the dominant effect of retrieval can be modeled by inserting a beamsplitter at the output to model photon loss.  Such a beamsplitter does not affect the pulse-integrated $g^{(2)}(0)$. This assumption is equivalent to stating that the probability of retrieving two photons from a doubly excited spin wave is $p_r^2$, where $p_r$ is the probability of retrieving a single photon from a singly excited spin-wave. However, due to blockade, multiple excitations occur predominantly near the edges of the ensemble \cite{Dudin.2012sx}, where the local atomic density is lower than in the center of the cloud. Since the retrieval efficiency of mode-matched photons increases with optical depth, the spatial dependence of the wavefunctions may alter the $g^{(2)}(0)$ in ways that our model does not take into account.

\section{Effect of Rabi frequency and hold time on $g^{(2)}(0)$}\label{sec:Dephasing}

\begin{figure}[t!]
\centering
\includegraphics{"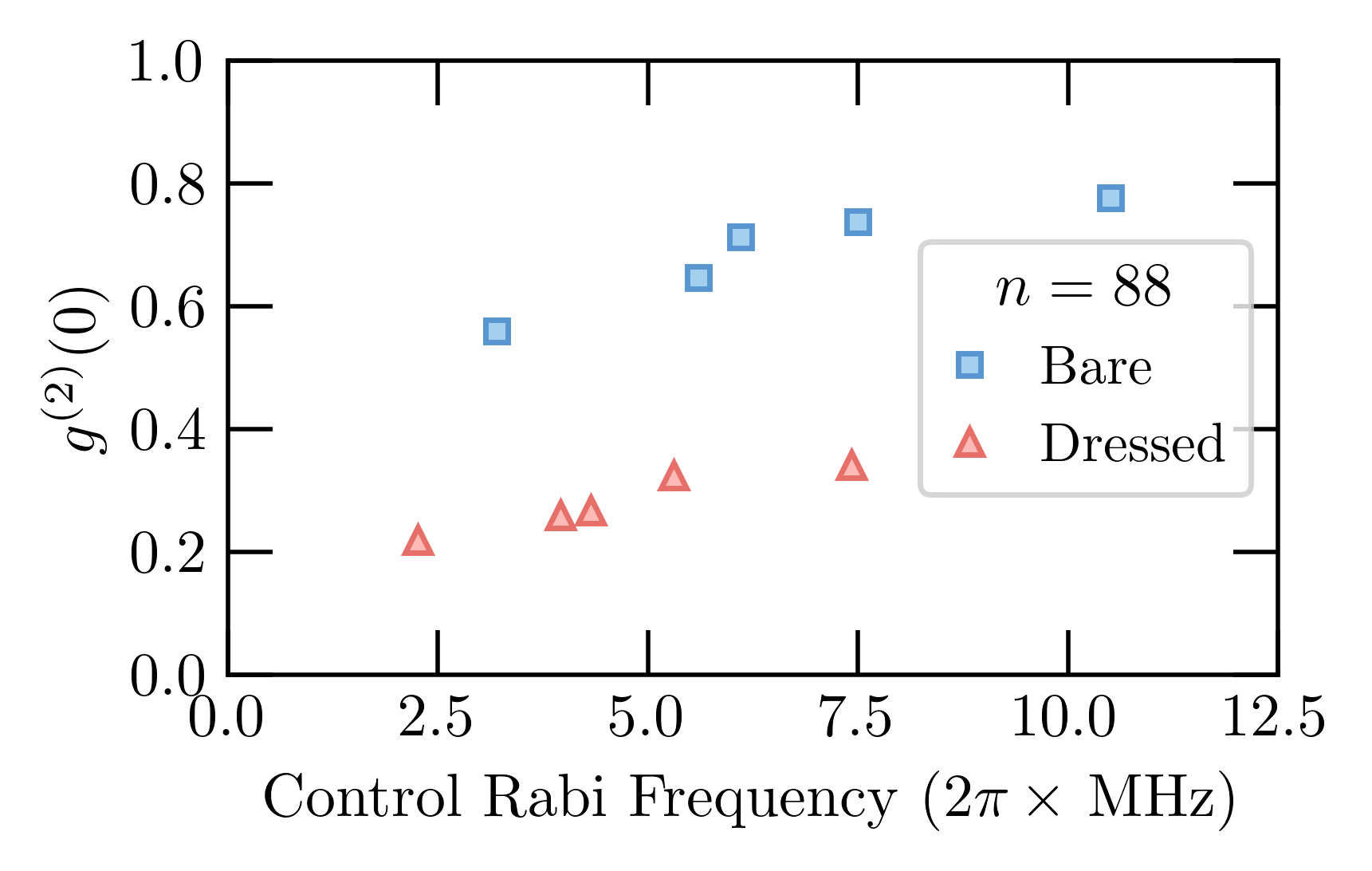"}
\caption{\label{fig:SupFig6} The impact of the control Rabi frequency on $g^{(2)}(0)$. The blue squares and red triangles correspond to data collected using, respectively, bare and dressed eigenstates at $n=88$.
$\Omega_c$ for bare states was measured through EIT spectroscopy, and the corresponding $\Omega_c$ for the dressed states was calculated by scaling the bare-state $\Omega_c$ by a factor of $1/\sqrt{2}$. The error bars, smaller than the markers used, indicate the $\pm \sigma$ statistical uncertainties. }
\end{figure}

In this section, we explore alternative mechanisms through which microwave dressing may reduce the $g^{(2)}(0)$. One potential cause for the reduction of $g^{(2)}(0)$ with microwave dressing is due to the relationship between the blockade radius and the Rydberg excitation Rabi frequency. On resonance, the $\ket{-}$ state is equally composed of $nS_{1/2}$ and $nP_{3/2}$, where the latter state does not couple to $\ket{e}$. Therefore, dressing leads to the reduction of $\Omega_c$ and correspondingly $\Omega_{\mathrm{eff},s}$ by $\sqrt{2}$. 
To explore the impact of $\Omega_c$ on single-photon purity, we reduce the control power and measure the corresponding reduction in $g^{(2)}(0)$ shown in Fig. \ref{fig:SupFig6}.
Comparing the $g^{(2)}(0)$ of the light produced using bare and dressed states at the same Rabi frequency excludes dressing-induced suppression of $g^{(2)}(0)$ by this mechanism.

Another possible cause for the suppression of $g^{(2)}(0)$ is the enhancement of interaction-induced dephasing of multiply excited spin-waves during storage \cite{Bariani.2011,Bariani.2012}.
This was explored in previous experiments where either exciting spin-waves in weakly interacting bare states \cite{Maxwell.2013,Maxwell.2014} or performing spectrally broad EIT storage sequences \cite{Xu.2024} resulted in poorly blockaded ensembles.
The multi-excitation components were then allowed to dephase over an extended storage duration yielding light with low $g^{(2)}(0)$.
In Refs.~\cite{Maxwell.2013,Maxwell.2014}, the application of microwave pulses accelerated multi-excitation dephasing through the enhancement of the interactions above those present in the excitation stage. 
In Ref.~\cite{Xu.2024}, the spin-waves were stored for much longer than the length of the Rydberg excitation pulse, allowing for the $1/r^3$ tails of the microwave-dressed-pair interactions to accelerate the dephasing of the distant pairs for which the $1/r^6$ bare interactions were vanishingly small.

\begin{figure}[b!]
    \centering
    \includegraphics[width=0.75\textwidth]{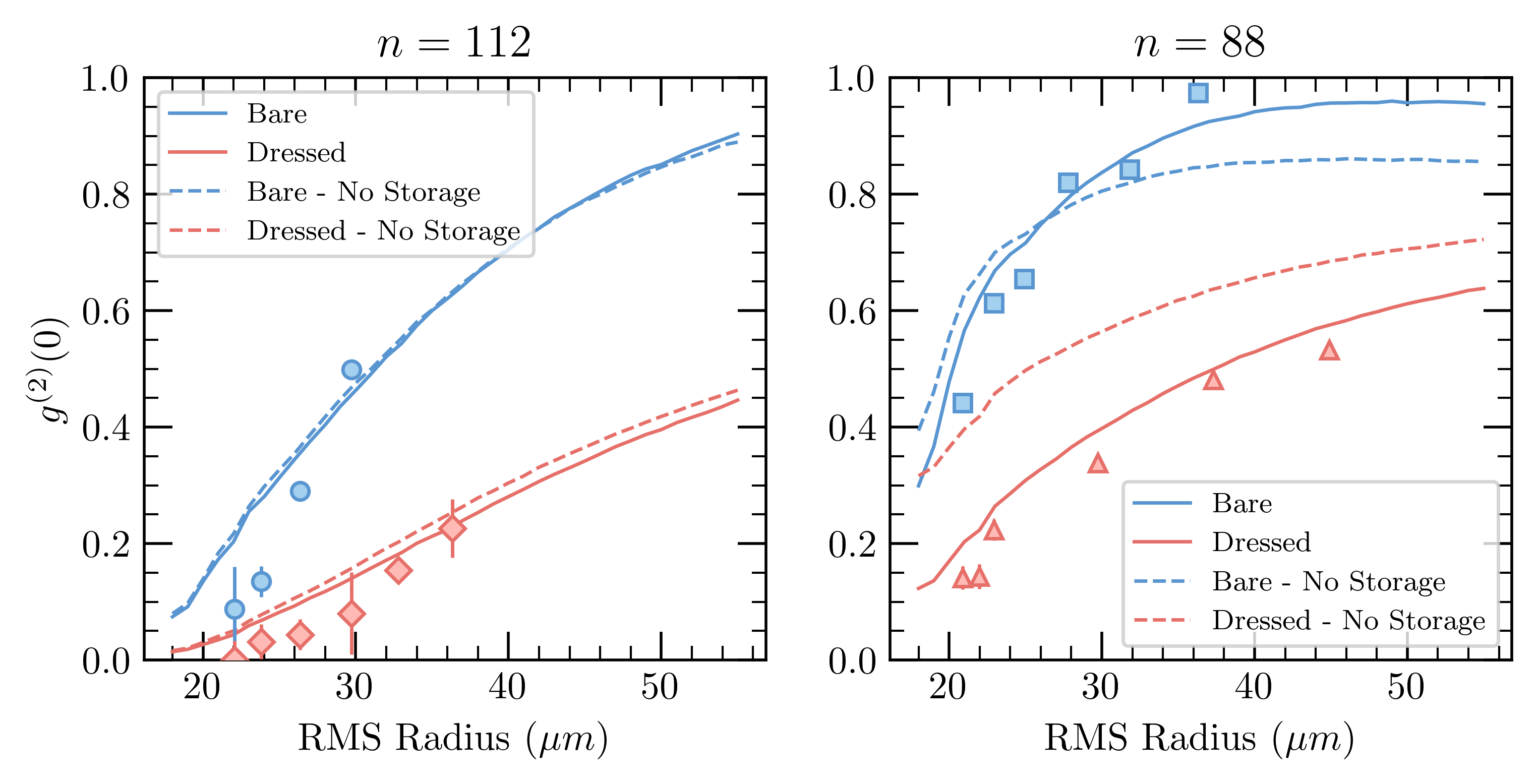}
    \caption{Calculated effect of the excitation storage time on the observed $g^{(2)}(0)$. We show theoretical calculations for $g^{(2)}(0)$ that do not include the $200$ ns storage time used in the experimental sequence (dashed curves), with results for $n=112$ (left) and $n=88$ (right). We also show the experimental (symbols) and theoretical (solid lines) results from Fig.~2(c) in the main text, for comparison.}
    \label{fig:SupFig7}
\end{figure}

Here, we use the theoretical model described in Sec.~\ref{sec:theory_model} to analyze the effect of the hold time on $g^{(2)}(0)$ in the experiment. In Fig.~\ref{fig:SupFig7}, we show the calculated $g^{(2)}(0)$ when the excitations are immediately read out following the write sequence, compared to both the experimental data and theoretical calculations from Fig.~2(c,d) of the main text, which include a $200$ ns hold time in between the write and retrieve stages. During this time, the relevant spin wave excitations can undergo interaction-induced dephasing, leading to a smaller observed value of $g^{(2)}$.

For $n=112$, we find that this storage time has little effect on the resulting $g^{(2)}(0)$. This is consistent with the relatively small two-photon Rabi frequency used during the write stage, which only enables the creation of excitations pairs whose interaction energy is either smaller than or comparable to this energy scale. As the total write time is $\sim1 \mu$s, we expect that interaction-induced dephasing will only occur over timescales much longer than the 200 ns storage time.

In contrast, we find that the storage time has a relatively noticeable effect for the $n=88$ results. This is consistent with the fact that the much faster $\sim 200$ ns write time is comparable to the storage time, so that interaction-induced dephasing effects may contribute significantly while the excitations are held in the spin wave states. We note that even for zero hold time simulations, microwave dressing substantially reduces $g^{(2)}(0)$, indicating that dephasing during the storage period alone cannot account of the enhancement of the single photon purity. For small values of $g^{(2)}(0)$, where $p_1 \gg p_2 \gg p_{n>2}$, interaction effects during the storage time will lead to a reduction in $p_2$, as the relevant spin wave excitation dephases. This leads to a net reduction in the value of $g^{(2)}(0)$. We note that, for larger values of $g^{(2)}(0)$, such as for large RMS cloud radii with the bare $n=88$ state interactions, our calculations suggest that interaction-induced dephasing can actually enhance the value of $g^{(2)}(0)$. However, for larger clouds and smaller blockade radii, it becomes necessary to examine the effect of more excitations in the cloud than we include in our calculations.


\begin{thebibliography}{65}%
\makeatletter
\providecommand \@ifxundefined [1]{%
 \@ifx{#1\undefined}
}%
\providecommand \@ifnum [1]{%
 \ifnum #1\expandafter \@firstoftwo
 \else \expandafter \@secondoftwo
 \fi
}%
\providecommand \@ifx [1]{%
 \ifx #1\expandafter \@firstoftwo
 \else \expandafter \@secondoftwo
 \fi
}%
\providecommand \natexlab [1]{#1}%
\providecommand \enquote  [1]{``#1''}%
\providecommand \bibnamefont  [1]{#1}%
\providecommand \bibfnamefont [1]{#1}%
\providecommand \citenamefont [1]{#1}%
\providecommand \href@noop [0]{\@secondoftwo}%
\providecommand \href [0]{\begingroup \@sanitize@url \@href}%
\providecommand \@href[1]{\@@startlink{#1}\@@href}%
\providecommand \@@href[1]{\endgroup#1\@@endlink}%
\providecommand \@sanitize@url [0]{\catcode `\\12\catcode `\$12\catcode
  `\&12\catcode `\#12\catcode `\^12\catcode `\_12\catcode `\%12\relax}%
\providecommand \@@startlink[1]{}%
\providecommand \@@endlink[0]{}%
\providecommand \url  [0]{\begingroup\@sanitize@url \@url }%
\providecommand \@url [1]{\endgroup\@href {#1}{\urlprefix }}%
\providecommand \urlprefix  [0]{URL }%
\providecommand \Eprint [0]{\href }%
\providecommand \doibase [0]{https://doi.org/}%
\providecommand \selectlanguage [0]{\@gobble}%
\providecommand \bibinfo  [0]{\@secondoftwo}%
\providecommand \bibfield  [0]{\@secondoftwo}%
\providecommand \translation [1]{[#1]}%
\providecommand \BibitemOpen [0]{}%
\providecommand \bibitemStop [0]{}%
\providecommand \bibitemNoStop [0]{.\EOS\space}%
\providecommand \EOS [0]{\spacefactor3000\relax}%
\providecommand \BibitemShut  [1]{\csname bibitem#1\endcsname}%
\let\auto@bib@innerbib\@empty
\bibitem [{\citenamefont {Saffman}\ \emph {et~al.}(2010)\citenamefont
  {Saffman}, \citenamefont {Walker},\ and\ \citenamefont
  {Mølmer}}]{Saffman.2010}%
  \BibitemOpen
  \bibfield  {author} {\bibinfo {author} {\bibfnamefont {M.}~\bibnamefont
  {Saffman}}, \bibinfo {author} {\bibfnamefont {T.~G.}\ \bibnamefont
  {Walker}},\ and\ \bibinfo {author} {\bibfnamefont {K.}~\bibnamefont
  {Mølmer}},\ }\bibfield  {title} {\bibinfo {title} {{Quantum information with
  Rydberg atoms}},\ }\href {https://doi.org/10.1103/revmodphys.82.2313}
  {\bibfield  {journal} {\bibinfo  {journal} {Reviews of Modern Physics}\
  }\textbf {\bibinfo {volume} {82}},\ \bibinfo {pages} {2313} (\bibinfo {year}
  {2010})},\ \Eprint {https://arxiv.org/abs/0909.4777} {0909.4777} \BibitemShut
  {NoStop}%
\bibitem [{\citenamefont {Browaeys}\ and\ \citenamefont
  {Lahaye}(2020)}]{Browaeys.2020}%
  \BibitemOpen
  \bibfield  {author} {\bibinfo {author} {\bibfnamefont {A.}~\bibnamefont
  {Browaeys}}\ and\ \bibinfo {author} {\bibfnamefont {T.}~\bibnamefont
  {Lahaye}},\ }\bibfield  {title} {\bibinfo {title} {{Many-body physics with
  individually controlled Rydberg atoms}},\ }\href
  {https://doi.org/10.1038/s41567-019-0733-z} {\bibfield  {journal} {\bibinfo
  {journal} {Nature Physics}\ }\textbf {\bibinfo {volume} {16}},\ \bibinfo
  {pages} {132} (\bibinfo {year} {2020})},\ \Eprint
  {https://arxiv.org/abs/2002.07413} {2002.07413} \BibitemShut {NoStop}%
\bibitem [{\citenamefont {Fan}\ \emph {et~al.}(2015)\citenamefont {Fan},
  \citenamefont {Kumar}, \citenamefont {Sedlacek}, \citenamefont {Kübler},
  \citenamefont {Karimkashi},\ and\ \citenamefont {Shaffer}}]{Fan.2015}%
  \BibitemOpen
  \bibfield  {author} {\bibinfo {author} {\bibfnamefont {H.}~\bibnamefont
  {Fan}}, \bibinfo {author} {\bibfnamefont {S.}~\bibnamefont {Kumar}}, \bibinfo
  {author} {\bibfnamefont {J.}~\bibnamefont {Sedlacek}}, \bibinfo {author}
  {\bibfnamefont {H.}~\bibnamefont {Kübler}}, \bibinfo {author} {\bibfnamefont
  {S.}~\bibnamefont {Karimkashi}},\ and\ \bibinfo {author} {\bibfnamefont
  {J.~P.}\ \bibnamefont {Shaffer}},\ }\bibfield  {title} {\bibinfo {title}
  {{Atom based RF electric field sensing}},\ }\href
  {https://doi.org/10.1088/0953-4075/48/20/202001} {\bibfield  {journal}
  {\bibinfo  {journal} {Journal of Physics B: Atomic, Molecular and Optical
  Physics}\ }\textbf {\bibinfo {volume} {48}},\ \bibinfo {pages} {202001}
  (\bibinfo {year} {2015})}\BibitemShut {NoStop}%
\bibitem [{\citenamefont {Firstenberg}\ \emph {et~al.}(2016)\citenamefont
  {Firstenberg}, \citenamefont {Adams},\ and\ \citenamefont
  {Hofferberth}}]{Firstenberg.2016}%
  \BibitemOpen
  \bibfield  {author} {\bibinfo {author} {\bibfnamefont {O.}~\bibnamefont
  {Firstenberg}}, \bibinfo {author} {\bibfnamefont {C.~S.}\ \bibnamefont
  {Adams}},\ and\ \bibinfo {author} {\bibfnamefont {S.}~\bibnamefont
  {Hofferberth}},\ }\bibfield  {title} {\bibinfo {title} {{Nonlinear quantum
  optics mediated by Rydberg interactions}},\ }\href
  {https://doi.org/10.1088/0953-4075/49/15/152003} {\bibfield  {journal}
  {\bibinfo  {journal} {Journal of Physics B: Atomic, Molecular and Optical
  Physics}\ }\textbf {\bibinfo {volume} {49}},\ \bibinfo {pages} {152003}
  (\bibinfo {year} {2016})},\ \Eprint {https://arxiv.org/abs/1602.06117}
  {1602.06117} \BibitemShut {NoStop}%
\bibitem [{\citenamefont {Gallagher}(1994)}]{Gallagher.1994}%
  \BibitemOpen
  \bibfield  {author} {\bibinfo {author} {\bibfnamefont {T.~F.}\ \bibnamefont
  {Gallagher}},\ }\href {https://doi.org/10.1017/cbo9780511524530} {\emph
  {\bibinfo {title} {{Rydberg Atoms}}}}\ (\bibinfo  {publisher} {Cambridge
  University Press},\ \bibinfo {year} {1994})\BibitemShut {NoStop}%
\bibitem [{\citenamefont {Jaksch}\ \emph {et~al.}(2000)\citenamefont {Jaksch},
  \citenamefont {Cirac}, \citenamefont {Zoller}, \citenamefont {Rolston},
  \citenamefont {Côté},\ and\ \citenamefont {Lukin}}]{Jaksch.2000}%
  \BibitemOpen
  \bibfield  {author} {\bibinfo {author} {\bibfnamefont {D.}~\bibnamefont
  {Jaksch}}, \bibinfo {author} {\bibfnamefont {J.~I.}\ \bibnamefont {Cirac}},
  \bibinfo {author} {\bibfnamefont {P.}~\bibnamefont {Zoller}}, \bibinfo
  {author} {\bibfnamefont {S.~L.}\ \bibnamefont {Rolston}}, \bibinfo {author}
  {\bibfnamefont {R.}~\bibnamefont {Côté}},\ and\ \bibinfo {author}
  {\bibfnamefont {M.~D.}\ \bibnamefont {Lukin}},\ }\bibfield  {title} {\bibinfo
  {title} {{Fast Quantum Gates for Neutral Atoms}},\ }\href
  {https://doi.org/10.1103/physrevlett.85.2208} {\bibfield  {journal} {\bibinfo
   {journal} {Physical Review Letters}\ }\textbf {\bibinfo {volume} {85}},\
  \bibinfo {pages} {2208} (\bibinfo {year} {2000})},\ \Eprint
  {https://arxiv.org/abs/quant-ph/0004038} {quant-ph/0004038} \BibitemShut
  {NoStop}%
\bibitem [{\citenamefont {Lukin}\ \emph {et~al.}(2000)\citenamefont {Lukin},
  \citenamefont {Fleischhauer}, \citenamefont {Cote}, \citenamefont {Duan},
  \citenamefont {Jaksch}, \citenamefont {Cirac},\ and\ \citenamefont
  {Zoller}}]{Lukin.2000}%
  \BibitemOpen
  \bibfield  {author} {\bibinfo {author} {\bibfnamefont {M.~D.}\ \bibnamefont
  {Lukin}}, \bibinfo {author} {\bibfnamefont {M.}~\bibnamefont {Fleischhauer}},
  \bibinfo {author} {\bibfnamefont {R.}~\bibnamefont {Cote}}, \bibinfo {author}
  {\bibfnamefont {L.~M.}\ \bibnamefont {Duan}}, \bibinfo {author}
  {\bibfnamefont {D.}~\bibnamefont {Jaksch}}, \bibinfo {author} {\bibfnamefont
  {J.~I.}\ \bibnamefont {Cirac}},\ and\ \bibinfo {author} {\bibfnamefont
  {P.}~\bibnamefont {Zoller}},\ }\bibfield  {title} {\bibinfo {title} {{Dipole
  Blockade and Quantum Information Processing in Mesoscopic Atomic
  Ensembles}},\ }\href {https://doi.org/10.1103/physrevlett.87.037901}
  {\bibfield  {journal} {\bibinfo  {journal} {Physical Review Letters}\
  }\textbf {\bibinfo {volume} {87}},\ \bibinfo {pages} {037901} (\bibinfo
  {year} {2000})},\ \Eprint {https://arxiv.org/abs/quant-ph/0011028}
  {quant-ph/0011028} \BibitemShut {NoStop}%
\bibitem [{\citenamefont {Saffman}\ and\ \citenamefont
  {Walker}(2005)}]{Saffman.2005}%
  \BibitemOpen
  \bibfield  {author} {\bibinfo {author} {\bibfnamefont {M.}~\bibnamefont
  {Saffman}}\ and\ \bibinfo {author} {\bibfnamefont {T.~G.}\ \bibnamefont
  {Walker}},\ }\bibfield  {title} {\bibinfo {title} {{Analysis of a quantum
  logic device based on dipole-dipole interactions of optically trapped Rydberg
  atoms}},\ }\href {https://doi.org/10.1103/physreva.72.022347} {\bibfield
  {journal} {\bibinfo  {journal} {Physical Review A}\ }\textbf {\bibinfo
  {volume} {72}},\ \bibinfo {pages} {022347} (\bibinfo {year} {2005})},\
  \Eprint {https://arxiv.org/abs/quant-ph/0502051} {quant-ph/0502051}
  \BibitemShut {NoStop}%
\bibitem [{\citenamefont {Levine}\ \emph {et~al.}(2018)\citenamefont {Levine},
  \citenamefont {Keesling}, \citenamefont {Omran}, \citenamefont {Bernien},
  \citenamefont {Schwartz}, \citenamefont {Zibrov}, \citenamefont {Endres},
  \citenamefont {Greiner}, \citenamefont {Vuletić},\ and\ \citenamefont
  {Lukin}}]{Levine.2018}%
  \BibitemOpen
  \bibfield  {author} {\bibinfo {author} {\bibfnamefont {H.}~\bibnamefont
  {Levine}}, \bibinfo {author} {\bibfnamefont {A.}~\bibnamefont {Keesling}},
  \bibinfo {author} {\bibfnamefont {A.}~\bibnamefont {Omran}}, \bibinfo
  {author} {\bibfnamefont {H.}~\bibnamefont {Bernien}}, \bibinfo {author}
  {\bibfnamefont {S.}~\bibnamefont {Schwartz}}, \bibinfo {author}
  {\bibfnamefont {A.~S.}\ \bibnamefont {Zibrov}}, \bibinfo {author}
  {\bibfnamefont {M.}~\bibnamefont {Endres}}, \bibinfo {author} {\bibfnamefont
  {M.}~\bibnamefont {Greiner}}, \bibinfo {author} {\bibfnamefont
  {V.}~\bibnamefont {Vuletić}},\ and\ \bibinfo {author} {\bibfnamefont
  {M.~D.}\ \bibnamefont {Lukin}},\ }\bibfield  {title} {\bibinfo {title}
  {{High-Fidelity Control and Entanglement of Rydberg-Atom Qubits}},\ }\href
  {https://doi.org/10.1103/physrevlett.121.123603} {\bibfield  {journal}
  {\bibinfo  {journal} {Physical Review Letters}\ }\textbf {\bibinfo {volume}
  {121}},\ \bibinfo {pages} {123603} (\bibinfo {year} {2018})},\ \Eprint
  {https://arxiv.org/abs/1806.04682} {1806.04682} \BibitemShut {NoStop}%
\bibitem [{\citenamefont {Glaetzle}\ \emph {et~al.}(2015)\citenamefont
  {Glaetzle}, \citenamefont {Dalmonte}, \citenamefont {Nath}, \citenamefont
  {Gross}, \citenamefont {Bloch},\ and\ \citenamefont
  {Zoller}}]{Glaetzle.2015}%
  \BibitemOpen
  \bibfield  {author} {\bibinfo {author} {\bibfnamefont {A.~W.}\ \bibnamefont
  {Glaetzle}}, \bibinfo {author} {\bibfnamefont {M.}~\bibnamefont {Dalmonte}},
  \bibinfo {author} {\bibfnamefont {R.}~\bibnamefont {Nath}}, \bibinfo {author}
  {\bibfnamefont {C.}~\bibnamefont {Gross}}, \bibinfo {author} {\bibfnamefont
  {I.}~\bibnamefont {Bloch}},\ and\ \bibinfo {author} {\bibfnamefont
  {P.}~\bibnamefont {Zoller}},\ }\bibfield  {title} {\bibinfo {title}
  {{Designing Frustrated Quantum Magnets with Laser-Dressed Rydberg Atoms}},\
  }\href {https://doi.org/10.1103/physrevlett.114.173002} {\bibfield  {journal}
  {\bibinfo  {journal} {Physical Review Letters}\ }\textbf {\bibinfo {volume}
  {114}},\ \bibinfo {pages} {173002} (\bibinfo {year} {2015})},\ \Eprint
  {https://arxiv.org/abs/1410.3388} {1410.3388} \BibitemShut {NoStop}%
\bibitem [{\citenamefont {Barredo}\ \emph {et~al.}(2015)\citenamefont
  {Barredo}, \citenamefont {Labuhn}, \citenamefont {Ravets}, \citenamefont
  {Lahaye}, \citenamefont {Browaeys},\ and\ \citenamefont
  {Adams}}]{Barredo.2015}%
  \BibitemOpen
  \bibfield  {author} {\bibinfo {author} {\bibfnamefont {D.}~\bibnamefont
  {Barredo}}, \bibinfo {author} {\bibfnamefont {H.}~\bibnamefont {Labuhn}},
  \bibinfo {author} {\bibfnamefont {S.}~\bibnamefont {Ravets}}, \bibinfo
  {author} {\bibfnamefont {T.}~\bibnamefont {Lahaye}}, \bibinfo {author}
  {\bibfnamefont {A.}~\bibnamefont {Browaeys}},\ and\ \bibinfo {author}
  {\bibfnamefont {C.~S.}\ \bibnamefont {Adams}},\ }\bibfield  {title} {\bibinfo
  {title} {{Coherent Excitation Transfer in a Spin Chain of Three Rydberg
  Atoms}},\ }\href {https://doi.org/10.1103/physrevlett.114.113002} {\bibfield
  {journal} {\bibinfo  {journal} {Physical Review Letters}\ }\textbf {\bibinfo
  {volume} {114}},\ \bibinfo {pages} {113002} (\bibinfo {year} {2015})},\
  \Eprint {https://arxiv.org/abs/1408.1055} {1408.1055} \BibitemShut {NoStop}%
\bibitem [{\citenamefont {Hollerith}\ \emph {et~al.}(2022)\citenamefont
  {Hollerith}, \citenamefont {Srakaew}, \citenamefont {Wei}, \citenamefont
  {Rubio-Abadal}, \citenamefont {Adler}, \citenamefont {Weckesser},
  \citenamefont {Kruckenhauser}, \citenamefont {Walther}, \citenamefont
  {Bijnen}, \citenamefont {Rui}, \citenamefont {Gross}, \citenamefont {Bloch},\
  and\ \citenamefont {Zeiher}}]{Hollerith.2022}%
  \BibitemOpen
  \bibfield  {author} {\bibinfo {author} {\bibfnamefont {S.}~\bibnamefont
  {Hollerith}}, \bibinfo {author} {\bibfnamefont {K.}~\bibnamefont {Srakaew}},
  \bibinfo {author} {\bibfnamefont {D.}~\bibnamefont {Wei}}, \bibinfo {author}
  {\bibfnamefont {A.}~\bibnamefont {Rubio-Abadal}}, \bibinfo {author}
  {\bibfnamefont {D.}~\bibnamefont {Adler}}, \bibinfo {author} {\bibfnamefont
  {P.}~\bibnamefont {Weckesser}}, \bibinfo {author} {\bibfnamefont
  {A.}~\bibnamefont {Kruckenhauser}}, \bibinfo {author} {\bibfnamefont
  {V.}~\bibnamefont {Walther}}, \bibinfo {author} {\bibfnamefont {R.~v.}\
  \bibnamefont {Bijnen}}, \bibinfo {author} {\bibfnamefont {J.}~\bibnamefont
  {Rui}}, \bibinfo {author} {\bibfnamefont {C.}~\bibnamefont {Gross}}, \bibinfo
  {author} {\bibfnamefont {I.}~\bibnamefont {Bloch}},\ and\ \bibinfo {author}
  {\bibfnamefont {J.}~\bibnamefont {Zeiher}},\ }\bibfield  {title} {\bibinfo
  {title} {{Realizing Distance-Selective Interactions in a Rydberg-Dressed Atom
  Array}},\ }\href {https://doi.org/10.1103/physrevlett.128.113602} {\bibfield
  {journal} {\bibinfo  {journal} {Physical Review Letters}\ }\textbf {\bibinfo
  {volume} {128}},\ \bibinfo {pages} {113602} (\bibinfo {year} {2022})},\
  \Eprint {https://arxiv.org/abs/2110.10125} {2110.10125} \BibitemShut
  {NoStop}%
\bibitem [{\citenamefont {Scholl}\ \emph {et~al.}(2022)\citenamefont {Scholl},
  \citenamefont {Williams}, \citenamefont {Bornet}, \citenamefont {Wallner},
  \citenamefont {Barredo}, \citenamefont {Henriet}, \citenamefont {Signoles},
  \citenamefont {Hainaut}, \citenamefont {Franz}, \citenamefont {Geier},
  \citenamefont {Tebben}, \citenamefont {Salzinger}, \citenamefont {Zürn},
  \citenamefont {Lahaye}, \citenamefont {Weidemüller},\ and\ \citenamefont
  {Browaeys}}]{Scholl.2022}%
  \BibitemOpen
  \bibfield  {author} {\bibinfo {author} {\bibfnamefont {P.}~\bibnamefont
  {Scholl}}, \bibinfo {author} {\bibfnamefont {H.~J.}\ \bibnamefont
  {Williams}}, \bibinfo {author} {\bibfnamefont {G.}~\bibnamefont {Bornet}},
  \bibinfo {author} {\bibfnamefont {F.}~\bibnamefont {Wallner}}, \bibinfo
  {author} {\bibfnamefont {D.}~\bibnamefont {Barredo}}, \bibinfo {author}
  {\bibfnamefont {L.}~\bibnamefont {Henriet}}, \bibinfo {author} {\bibfnamefont
  {A.}~\bibnamefont {Signoles}}, \bibinfo {author} {\bibfnamefont
  {C.}~\bibnamefont {Hainaut}}, \bibinfo {author} {\bibfnamefont
  {T.}~\bibnamefont {Franz}}, \bibinfo {author} {\bibfnamefont
  {S.}~\bibnamefont {Geier}}, \bibinfo {author} {\bibfnamefont
  {A.}~\bibnamefont {Tebben}}, \bibinfo {author} {\bibfnamefont
  {A.}~\bibnamefont {Salzinger}}, \bibinfo {author} {\bibfnamefont
  {G.}~\bibnamefont {Zürn}}, \bibinfo {author} {\bibfnamefont
  {T.}~\bibnamefont {Lahaye}}, \bibinfo {author} {\bibfnamefont
  {M.}~\bibnamefont {Weidemüller}},\ and\ \bibinfo {author} {\bibfnamefont
  {A.}~\bibnamefont {Browaeys}},\ }\bibfield  {title} {\bibinfo {title}
  {{Microwave Engineering of Programmable XXZ Hamiltonians in Arrays of Rydberg
  Atoms}},\ }\href {https://doi.org/10.1103/prxquantum.3.020303} {\bibfield
  {journal} {\bibinfo  {journal} {PRX Quantum}\ }\textbf {\bibinfo {volume}
  {3}},\ \bibinfo {pages} {020303} (\bibinfo {year} {2022})},\ \Eprint
  {https://arxiv.org/abs/2107.14459} {2107.14459} \BibitemShut {NoStop}%
\bibitem [{\citenamefont {Steinert}\ \emph {et~al.}(2023)\citenamefont
  {Steinert}, \citenamefont {Osterholz}, \citenamefont {Eberhard},
  \citenamefont {Festa}, \citenamefont {Lorenz}, \citenamefont {Chen},
  \citenamefont {Trautmann},\ and\ \citenamefont {Gross}}]{Steinert.2023}%
  \BibitemOpen
  \bibfield  {author} {\bibinfo {author} {\bibfnamefont {L.-M.}\ \bibnamefont
  {Steinert}}, \bibinfo {author} {\bibfnamefont {P.}~\bibnamefont {Osterholz}},
  \bibinfo {author} {\bibfnamefont {R.}~\bibnamefont {Eberhard}}, \bibinfo
  {author} {\bibfnamefont {L.}~\bibnamefont {Festa}}, \bibinfo {author}
  {\bibfnamefont {N.}~\bibnamefont {Lorenz}}, \bibinfo {author} {\bibfnamefont
  {Z.}~\bibnamefont {Chen}}, \bibinfo {author} {\bibfnamefont {A.}~\bibnamefont
  {Trautmann}},\ and\ \bibinfo {author} {\bibfnamefont {C.}~\bibnamefont
  {Gross}},\ }\bibfield  {title} {\bibinfo {title} {{Spatially Tunable Spin
  Interactions in Neutral Atom Arrays}},\ }\href
  {https://doi.org/10.1103/physrevlett.130.243001} {\bibfield  {journal}
  {\bibinfo  {journal} {Physical Review Letters}\ }\textbf {\bibinfo {volume}
  {130}},\ \bibinfo {pages} {243001} (\bibinfo {year} {2023})},\ \Eprint
  {https://arxiv.org/abs/2206.12385} {2206.12385} \BibitemShut {NoStop}%
\bibitem [{\citenamefont {Signoles}\ \emph {et~al.}(2021)\citenamefont
  {Signoles}, \citenamefont {Franz}, \citenamefont {Alves}, \citenamefont
  {Gärttner}, \citenamefont {Whitlock}, \citenamefont {Zürn},\ and\
  \citenamefont {Weidemüller}}]{Signoles.2021}%
  \BibitemOpen
  \bibfield  {author} {\bibinfo {author} {\bibfnamefont {A.}~\bibnamefont
  {Signoles}}, \bibinfo {author} {\bibfnamefont {T.}~\bibnamefont {Franz}},
  \bibinfo {author} {\bibfnamefont {R.~F.}\ \bibnamefont {Alves}}, \bibinfo
  {author} {\bibfnamefont {M.}~\bibnamefont {Gärttner}}, \bibinfo {author}
  {\bibfnamefont {S.}~\bibnamefont {Whitlock}}, \bibinfo {author}
  {\bibfnamefont {G.}~\bibnamefont {Zürn}},\ and\ \bibinfo {author}
  {\bibfnamefont {M.}~\bibnamefont {Weidemüller}},\ }\bibfield  {title}
  {\bibinfo {title} {{Glassy Dynamics in a Disordered Heisenberg Quantum Spin
  System}},\ }\href {https://doi.org/10.1103/physrevx.11.011011} {\bibfield
  {journal} {\bibinfo  {journal} {Physical Review X}\ }\textbf {\bibinfo
  {volume} {11}},\ \bibinfo {pages} {011011} (\bibinfo {year} {2021})},\
  \Eprint {https://arxiv.org/abs/1909.11959} {1909.11959} \BibitemShut
  {NoStop}%
\bibitem [{\citenamefont {Zeiher}\ \emph {et~al.}(2016)\citenamefont {Zeiher},
  \citenamefont {Bijnen}, \citenamefont {Schauß}, \citenamefont {Hild},
  \citenamefont {Choi}, \citenamefont {Pohl}, \citenamefont {Bloch},\ and\
  \citenamefont {Gross}}]{Zeiher.2016}%
  \BibitemOpen
  \bibfield  {author} {\bibinfo {author} {\bibfnamefont {J.}~\bibnamefont
  {Zeiher}}, \bibinfo {author} {\bibfnamefont {R.~v.}\ \bibnamefont {Bijnen}},
  \bibinfo {author} {\bibfnamefont {P.}~\bibnamefont {Schauß}}, \bibinfo
  {author} {\bibfnamefont {S.}~\bibnamefont {Hild}}, \bibinfo {author}
  {\bibfnamefont {J.-y.}\ \bibnamefont {Choi}}, \bibinfo {author}
  {\bibfnamefont {T.}~\bibnamefont {Pohl}}, \bibinfo {author} {\bibfnamefont
  {I.}~\bibnamefont {Bloch}},\ and\ \bibinfo {author} {\bibfnamefont
  {C.}~\bibnamefont {Gross}},\ }\bibfield  {title} {\bibinfo {title}
  {{Many-body interferometry of a Rydberg-dressed spin lattice}},\ }\href
  {https://doi.org/10.1038/nphys3835} {\bibfield  {journal} {\bibinfo
  {journal} {Nature Physics}\ }\textbf {\bibinfo {volume} {12}},\ \bibinfo
  {pages} {1095} (\bibinfo {year} {2016})},\ \Eprint
  {https://arxiv.org/abs/1602.06313} {1602.06313} \BibitemShut {NoStop}%
\bibitem [{\citenamefont {Borish}\ \emph {et~al.}(2020)\citenamefont {Borish},
  \citenamefont {Marković}, \citenamefont {Hines}, \citenamefont {Rajagopal},\
  and\ \citenamefont {Schleier-Smith}}]{Borish.2020}%
  \BibitemOpen
  \bibfield  {author} {\bibinfo {author} {\bibfnamefont {V.}~\bibnamefont
  {Borish}}, \bibinfo {author} {\bibfnamefont {O.}~\bibnamefont {Marković}},
  \bibinfo {author} {\bibfnamefont {J.~A.}\ \bibnamefont {Hines}}, \bibinfo
  {author} {\bibfnamefont {S.~V.}\ \bibnamefont {Rajagopal}},\ and\ \bibinfo
  {author} {\bibfnamefont {M.}~\bibnamefont {Schleier-Smith}},\ }\bibfield
  {title} {\bibinfo {title} {{Transverse-Field Ising Dynamics in a
  Rydberg-Dressed Atomic Gas}},\ }\href
  {https://doi.org/10.1103/physrevlett.124.063601} {\bibfield  {journal}
  {\bibinfo  {journal} {Physical Review Letters}\ }\textbf {\bibinfo {volume}
  {124}},\ \bibinfo {pages} {063601} (\bibinfo {year} {2020})},\ \Eprint
  {https://arxiv.org/abs/1910.13687} {1910.13687} \BibitemShut {NoStop}%
\bibitem [{\citenamefont {Geier}\ \emph {et~al.}(2021)\citenamefont {Geier},
  \citenamefont {Thaicharoen}, \citenamefont {Hainaut}, \citenamefont {Franz},
  \citenamefont {Salzinger}, \citenamefont {Tebben}, \citenamefont
  {Grimshandl}, \citenamefont {Zürn},\ and\ \citenamefont
  {Weidemüller}}]{Geier.2021}%
  \BibitemOpen
  \bibfield  {author} {\bibinfo {author} {\bibfnamefont {S.}~\bibnamefont
  {Geier}}, \bibinfo {author} {\bibfnamefont {N.}~\bibnamefont {Thaicharoen}},
  \bibinfo {author} {\bibfnamefont {C.}~\bibnamefont {Hainaut}}, \bibinfo
  {author} {\bibfnamefont {T.}~\bibnamefont {Franz}}, \bibinfo {author}
  {\bibfnamefont {A.}~\bibnamefont {Salzinger}}, \bibinfo {author}
  {\bibfnamefont {A.}~\bibnamefont {Tebben}}, \bibinfo {author} {\bibfnamefont
  {D.}~\bibnamefont {Grimshandl}}, \bibinfo {author} {\bibfnamefont
  {G.}~\bibnamefont {Zürn}},\ and\ \bibinfo {author} {\bibfnamefont
  {M.}~\bibnamefont {Weidemüller}},\ }\bibfield  {title} {\bibinfo {title}
  {{Floquet Hamiltonian engineering of an isolated many-body spin system}},\
  }\href {https://doi.org/10.1126/science.abd9547} {\bibfield  {journal}
  {\bibinfo  {journal} {Science}\ }\textbf {\bibinfo {volume} {374}},\ \bibinfo
  {pages} {1149} (\bibinfo {year} {2021})},\ \Eprint
  {https://arxiv.org/abs/2105.01597} {2105.01597} \BibitemShut {NoStop}%
\bibitem [{\citenamefont {Peyronel}\ \emph {et~al.}(2012)\citenamefont
  {Peyronel}, \citenamefont {Firstenberg}, \citenamefont {Liang}, \citenamefont
  {Hofferberth}, \citenamefont {Gorshkov}, \citenamefont {Pohl}, \citenamefont
  {Lukin},\ and\ \citenamefont {Vuletić}}]{Peyronel.2012}%
  \BibitemOpen
  \bibfield  {author} {\bibinfo {author} {\bibfnamefont {T.}~\bibnamefont
  {Peyronel}}, \bibinfo {author} {\bibfnamefont {O.}~\bibnamefont
  {Firstenberg}}, \bibinfo {author} {\bibfnamefont {Q.-Y.}\ \bibnamefont
  {Liang}}, \bibinfo {author} {\bibfnamefont {S.}~\bibnamefont {Hofferberth}},
  \bibinfo {author} {\bibfnamefont {A.~V.}\ \bibnamefont {Gorshkov}}, \bibinfo
  {author} {\bibfnamefont {T.}~\bibnamefont {Pohl}}, \bibinfo {author}
  {\bibfnamefont {M.~D.}\ \bibnamefont {Lukin}},\ and\ \bibinfo {author}
  {\bibfnamefont {V.}~\bibnamefont {Vuletić}},\ }\bibfield  {title} {\bibinfo
  {title} {{Quantum nonlinear optics with single photons enabled by strongly
  interacting atoms}},\ }\href {https://doi.org/10.1038/nature11361} {\bibfield
   {journal} {\bibinfo  {journal} {Nature}\ }\textbf {\bibinfo {volume}
  {488}},\ \bibinfo {pages} {57} (\bibinfo {year} {2012})}\BibitemShut
  {NoStop}%
\bibitem [{\citenamefont {Kanungo}\ \emph {et~al.}(2022)\citenamefont
  {Kanungo}, \citenamefont {Whalen}, \citenamefont {Lu}, \citenamefont {Yuan},
  \citenamefont {Dasgupta}, \citenamefont {Dunning}, \citenamefont {Hazzard},\
  and\ \citenamefont {Killian}}]{Kanungo.2022}%
  \BibitemOpen
  \bibfield  {author} {\bibinfo {author} {\bibfnamefont {S.~K.}\ \bibnamefont
  {Kanungo}}, \bibinfo {author} {\bibfnamefont {J.~D.}\ \bibnamefont {Whalen}},
  \bibinfo {author} {\bibfnamefont {Y.}~\bibnamefont {Lu}}, \bibinfo {author}
  {\bibfnamefont {M.}~\bibnamefont {Yuan}}, \bibinfo {author} {\bibfnamefont
  {S.}~\bibnamefont {Dasgupta}}, \bibinfo {author} {\bibfnamefont {F.~B.}\
  \bibnamefont {Dunning}}, \bibinfo {author} {\bibfnamefont {K.~R.~A.}\
  \bibnamefont {Hazzard}},\ and\ \bibinfo {author} {\bibfnamefont {T.~C.}\
  \bibnamefont {Killian}},\ }\bibfield  {title} {\bibinfo {title} {{Realizing
  topological edge states with Rydberg-atom synthetic dimensions}},\ }\href
  {https://doi.org/10.1038/s41467-022-28550-y} {\bibfield  {journal} {\bibinfo
  {journal} {Nature Communications}\ }\textbf {\bibinfo {volume} {13}},\
  \bibinfo {pages} {972} (\bibinfo {year} {2022})},\ \Eprint
  {https://arxiv.org/abs/2101.02871} {2101.02871} \BibitemShut {NoStop}%
\bibitem [{\citenamefont {Ornelas-Huerta}\ \emph {et~al.}(2020)\citenamefont
  {Ornelas-Huerta}, \citenamefont {Craddock}, \citenamefont {Goldschmidt},
  \citenamefont {Hachtel}, \citenamefont {Wang}, \citenamefont {Bienias},
  \citenamefont {Gorshkov}, \citenamefont {Rolston},\ and\ \citenamefont
  {Porto}}]{Ornelas-Huerta.2020}%
  \BibitemOpen
  \bibfield  {author} {\bibinfo {author} {\bibfnamefont {D.~P.}\ \bibnamefont
  {Ornelas-Huerta}}, \bibinfo {author} {\bibfnamefont {A.~N.}\ \bibnamefont
  {Craddock}}, \bibinfo {author} {\bibfnamefont {E.~A.}\ \bibnamefont
  {Goldschmidt}}, \bibinfo {author} {\bibfnamefont {A.~J.}\ \bibnamefont
  {Hachtel}}, \bibinfo {author} {\bibfnamefont {Y.}~\bibnamefont {Wang}},
  \bibinfo {author} {\bibfnamefont {P.}~\bibnamefont {Bienias}}, \bibinfo
  {author} {\bibfnamefont {A.~V.}\ \bibnamefont {Gorshkov}}, \bibinfo {author}
  {\bibfnamefont {S.~L.}\ \bibnamefont {Rolston}},\ and\ \bibinfo {author}
  {\bibfnamefont {J.~V.}\ \bibnamefont {Porto}},\ }\bibfield  {title} {\bibinfo
  {title} {{On-demand indistinguishable single photons from an efficient and
  pure source based on a Rydberg ensemble}},\ }\href
  {https://doi.org/10.1364/optica.391485} {\bibfield  {journal} {\bibinfo
  {journal} {Optica}\ }\textbf {\bibinfo {volume} {7}},\ \bibinfo {pages} {813}
  (\bibinfo {year} {2020})},\ \Eprint {https://arxiv.org/abs/2003.02202}
  {2003.02202} \BibitemShut {NoStop}%
\bibitem [{\citenamefont {Yang}\ \emph {et~al.}(2022)\citenamefont {Yang},
  \citenamefont {Li}, \citenamefont {Zhou}, \citenamefont {Jiang},
  \citenamefont {Bao},\ and\ \citenamefont {Pan}}]{Yang.2022}%
  \BibitemOpen
  \bibfield  {author} {\bibinfo {author} {\bibfnamefont {C.-W.}\ \bibnamefont
  {Yang}}, \bibinfo {author} {\bibfnamefont {J.}~\bibnamefont {Li}}, \bibinfo
  {author} {\bibfnamefont {M.-T.}\ \bibnamefont {Zhou}}, \bibinfo {author}
  {\bibfnamefont {X.}~\bibnamefont {Jiang}}, \bibinfo {author} {\bibfnamefont
  {X.-H.}\ \bibnamefont {Bao}},\ and\ \bibinfo {author} {\bibfnamefont {J.-W.}\
  \bibnamefont {Pan}},\ }\bibfield  {title} {\bibinfo {title} {{Deterministic
  measurement of a Rydberg superatom qubit via cavity-enhanced single-photon
  emission}},\ }\href {https://doi.org/10.1364/optica.461287} {\bibfield
  {journal} {\bibinfo  {journal} {Optica}\ }\textbf {\bibinfo {volume} {9}},\
  \bibinfo {pages} {853} (\bibinfo {year} {2022})}\BibitemShut {NoStop}%
\bibitem [{\citenamefont {Shi}\ \emph {et~al.}(2022)\citenamefont {Shi},
  \citenamefont {Xu}, \citenamefont {Zhang}, \citenamefont {Ye}, \citenamefont
  {Xiang}, \citenamefont {Liu}, \citenamefont {Wang}, \citenamefont {Su},\ and\
  \citenamefont {Li}}]{Shi.2022}%
  \BibitemOpen
  \bibfield  {author} {\bibinfo {author} {\bibfnamefont {S.}~\bibnamefont
  {Shi}}, \bibinfo {author} {\bibfnamefont {B.}~\bibnamefont {Xu}}, \bibinfo
  {author} {\bibfnamefont {K.}~\bibnamefont {Zhang}}, \bibinfo {author}
  {\bibfnamefont {G.-S.}\ \bibnamefont {Ye}}, \bibinfo {author} {\bibfnamefont
  {D.-S.}\ \bibnamefont {Xiang}}, \bibinfo {author} {\bibfnamefont
  {Y.}~\bibnamefont {Liu}}, \bibinfo {author} {\bibfnamefont {J.}~\bibnamefont
  {Wang}}, \bibinfo {author} {\bibfnamefont {D.}~\bibnamefont {Su}},\ and\
  \bibinfo {author} {\bibfnamefont {L.}~\bibnamefont {Li}},\ }\bibfield
  {title} {\bibinfo {title} {{High-fidelity photonic quantum logic gate based
  on near-optimal Rydberg single-photon source}},\ }\href
  {https://doi.org/10.1038/s41467-022-32083-9} {\bibfield  {journal} {\bibinfo
  {journal} {Nature Communications}\ }\textbf {\bibinfo {volume} {13}},\
  \bibinfo {pages} {4454} (\bibinfo {year} {2022})}\BibitemShut {NoStop}%
\bibitem [{\citenamefont {Ye}\ \emph {et~al.}(2023)\citenamefont {Ye},
  \citenamefont {Xu}, \citenamefont {Chang}, \citenamefont {Shi}, \citenamefont
  {Shi},\ and\ \citenamefont {Li}}]{Ye.2023}%
  \BibitemOpen
  \bibfield  {author} {\bibinfo {author} {\bibfnamefont {G.-S.}\ \bibnamefont
  {Ye}}, \bibinfo {author} {\bibfnamefont {B.}~\bibnamefont {Xu}}, \bibinfo
  {author} {\bibfnamefont {Y.}~\bibnamefont {Chang}}, \bibinfo {author}
  {\bibfnamefont {S.}~\bibnamefont {Shi}}, \bibinfo {author} {\bibfnamefont
  {T.}~\bibnamefont {Shi}},\ and\ \bibinfo {author} {\bibfnamefont
  {L.}~\bibnamefont {Li}},\ }\bibfield  {title} {\bibinfo {title} {{A photonic
  entanglement filter with Rydberg atoms}},\ }\href
  {https://doi.org/10.1038/s41566-023-01194-0} {\bibfield  {journal} {\bibinfo
  {journal} {Nature Photonics}\ }\textbf {\bibinfo {volume} {17}},\ \bibinfo
  {pages} {538} (\bibinfo {year} {2023})},\ \Eprint
  {https://arxiv.org/abs/2209.03107} {2209.03107} \BibitemShut {NoStop}%
\bibitem [{\citenamefont {Tiarks}\ \emph {et~al.}(2019)\citenamefont {Tiarks},
  \citenamefont {Schmidt-Eberle}, \citenamefont {Stolz}, \citenamefont
  {Rempe},\ and\ \citenamefont {Dürr}}]{Tiarks.2019}%
  \BibitemOpen
  \bibfield  {author} {\bibinfo {author} {\bibfnamefont {D.}~\bibnamefont
  {Tiarks}}, \bibinfo {author} {\bibfnamefont {S.}~\bibnamefont
  {Schmidt-Eberle}}, \bibinfo {author} {\bibfnamefont {T.}~\bibnamefont
  {Stolz}}, \bibinfo {author} {\bibfnamefont {G.}~\bibnamefont {Rempe}},\ and\
  \bibinfo {author} {\bibfnamefont {S.}~\bibnamefont {Dürr}},\ }\bibfield
  {title} {\bibinfo {title} {{A photon–photon quantum gate based on Rydberg
  interactions}},\ }\href {https://doi.org/10.1038/s41567-018-0313-7}
  {\bibfield  {journal} {\bibinfo  {journal} {Nature Physics}\ }\textbf
  {\bibinfo {volume} {15}},\ \bibinfo {pages} {124} (\bibinfo {year} {2019})},\
  \Eprint {https://arxiv.org/abs/1807.05795} {1807.05795} \BibitemShut
  {NoStop}%
\bibitem [{\citenamefont {Tiarks}\ \emph {et~al.}(2014)\citenamefont {Tiarks},
  \citenamefont {Baur}, \citenamefont {Schneider}, \citenamefont {Dürr},\ and\
  \citenamefont {Rempe}}]{Tiarks.2014}%
  \BibitemOpen
  \bibfield  {author} {\bibinfo {author} {\bibfnamefont {D.}~\bibnamefont
  {Tiarks}}, \bibinfo {author} {\bibfnamefont {S.}~\bibnamefont {Baur}},
  \bibinfo {author} {\bibfnamefont {K.}~\bibnamefont {Schneider}}, \bibinfo
  {author} {\bibfnamefont {S.}~\bibnamefont {Dürr}},\ and\ \bibinfo {author}
  {\bibfnamefont {G.}~\bibnamefont {Rempe}},\ }\bibfield  {title} {\bibinfo
  {title} {{Single-Photon Transistor Using a Förster Resonance}},\ }\href
  {https://doi.org/10.1103/physrevlett.113.053602} {\bibfield  {journal}
  {\bibinfo  {journal} {Physical Review Letters}\ }\textbf {\bibinfo {volume}
  {113}},\ \bibinfo {pages} {053602} (\bibinfo {year} {2014})},\ \Eprint
  {https://arxiv.org/abs/1404.3061} {1404.3061} \BibitemShut {NoStop}%
\bibitem [{\citenamefont {Gorniaczyk}\ \emph {et~al.}(2014)\citenamefont
  {Gorniaczyk}, \citenamefont {Tresp}, \citenamefont {Schmidt}, \citenamefont
  {Fedder},\ and\ \citenamefont {Hofferberth}}]{Gorniaczyk.2014}%
  \BibitemOpen
  \bibfield  {author} {\bibinfo {author} {\bibfnamefont {H.}~\bibnamefont
  {Gorniaczyk}}, \bibinfo {author} {\bibfnamefont {C.}~\bibnamefont {Tresp}},
  \bibinfo {author} {\bibfnamefont {J.}~\bibnamefont {Schmidt}}, \bibinfo
  {author} {\bibfnamefont {H.}~\bibnamefont {Fedder}},\ and\ \bibinfo {author}
  {\bibfnamefont {S.}~\bibnamefont {Hofferberth}},\ }\bibfield  {title}
  {\bibinfo {title} {{Single-Photon Transistor Mediated by Interstate Rydberg
  Interactions}},\ }\href {https://doi.org/10.1103/physrevlett.113.053601}
  {\bibfield  {journal} {\bibinfo  {journal} {Physical Review Letters}\
  }\textbf {\bibinfo {volume} {113}},\ \bibinfo {pages} {053601} (\bibinfo
  {year} {2014})},\ \Eprint {https://arxiv.org/abs/1404.2876} {1404.2876}
  \BibitemShut {NoStop}%
\bibitem [{\citenamefont {Gorshkov}\ \emph {et~al.}(2011)\citenamefont
  {Gorshkov}, \citenamefont {Otterbach}, \citenamefont {Fleischhauer},
  \citenamefont {Pohl},\ and\ \citenamefont {Lukin}}]{Gorshkov.2011}%
  \BibitemOpen
  \bibfield  {author} {\bibinfo {author} {\bibfnamefont {A.~V.}\ \bibnamefont
  {Gorshkov}}, \bibinfo {author} {\bibfnamefont {J.}~\bibnamefont {Otterbach}},
  \bibinfo {author} {\bibfnamefont {M.}~\bibnamefont {Fleischhauer}}, \bibinfo
  {author} {\bibfnamefont {T.}~\bibnamefont {Pohl}},\ and\ \bibinfo {author}
  {\bibfnamefont {M.~D.}\ \bibnamefont {Lukin}},\ }\bibfield  {title} {\bibinfo
  {title} {{Photon-Photon Interactions via Rydberg Blockade}},\ }\href
  {https://doi.org/10.1103/physrevlett.107.133602} {\bibfield  {journal}
  {\bibinfo  {journal} {Physical Review Letters}\ }\textbf {\bibinfo {volume}
  {107}},\ \bibinfo {pages} {133602} (\bibinfo {year} {2011})},\ \Eprint
  {https://arxiv.org/abs/1103.3700} {1103.3700} \BibitemShut {NoStop}%
\bibitem [{\citenamefont {Graß}\ \emph {et~al.}(2018)\citenamefont {Graß},
  \citenamefont {Bienias}, \citenamefont {Gullans}, \citenamefont {Lundgren},
  \citenamefont {Maciejko},\ and\ \citenamefont {Gorshkov}}]{Grass.2018}%
  \BibitemOpen
  \bibfield  {author} {\bibinfo {author} {\bibfnamefont {T.}~\bibnamefont
  {Graß}}, \bibinfo {author} {\bibfnamefont {P.}~\bibnamefont {Bienias}},
  \bibinfo {author} {\bibfnamefont {M.~J.}\ \bibnamefont {Gullans}}, \bibinfo
  {author} {\bibfnamefont {R.}~\bibnamefont {Lundgren}}, \bibinfo {author}
  {\bibfnamefont {J.}~\bibnamefont {Maciejko}},\ and\ \bibinfo {author}
  {\bibfnamefont {A.~V.}\ \bibnamefont {Gorshkov}},\ }\bibfield  {title}
  {\bibinfo {title} {{Fractional Quantum Hall Phases of Bosons with Tunable
  Interactions: From the Laughlin Liquid to a Fractional Wigner Crystal}},\
  }\href {https://doi.org/10.1103/physrevlett.121.253403} {\bibfield  {journal}
  {\bibinfo  {journal} {Physical Review Letters}\ }\textbf {\bibinfo {volume}
  {121}},\ \bibinfo {pages} {253403} (\bibinfo {year} {2018})},\ \Eprint
  {https://arxiv.org/abs/1809.04493} {1809.04493} \BibitemShut {NoStop}%
\bibitem [{\citenamefont {Gullans}\ \emph {et~al.}(2017)\citenamefont
  {Gullans}, \citenamefont {Diehl}, \citenamefont {Rittenhouse}, \citenamefont
  {Ruzic}, \citenamefont {D’Incao}, \citenamefont {Julienne}, \citenamefont
  {Gorshkov},\ and\ \citenamefont {Taylor}}]{Gullans.2017}%
  \BibitemOpen
  \bibfield  {author} {\bibinfo {author} {\bibfnamefont {M.~J.}\ \bibnamefont
  {Gullans}}, \bibinfo {author} {\bibfnamefont {S.}~\bibnamefont {Diehl}},
  \bibinfo {author} {\bibfnamefont {S.~T.}\ \bibnamefont {Rittenhouse}},
  \bibinfo {author} {\bibfnamefont {B.~P.}\ \bibnamefont {Ruzic}}, \bibinfo
  {author} {\bibfnamefont {J.~P.}\ \bibnamefont {D’Incao}}, \bibinfo {author}
  {\bibfnamefont {P.}~\bibnamefont {Julienne}}, \bibinfo {author}
  {\bibfnamefont {A.~V.}\ \bibnamefont {Gorshkov}},\ and\ \bibinfo {author}
  {\bibfnamefont {J.~M.}\ \bibnamefont {Taylor}},\ }\bibfield  {title}
  {\bibinfo {title} {{Efimov States of Strongly Interacting Photons}},\ }\href
  {https://doi.org/10.1103/physrevlett.119.233601} {\bibfield  {journal}
  {\bibinfo  {journal} {Physical Review Letters}\ }\textbf {\bibinfo {volume}
  {119}},\ \bibinfo {pages} {233601} (\bibinfo {year} {2017})},\ \Eprint
  {https://arxiv.org/abs/1709.01955} {1709.01955} \BibitemShut {NoStop}%
\bibitem [{\citenamefont {Paredes-Barato}\ and\ \citenamefont
  {Adams}(2013)}]{Paredes-Barato.2013}%
  \BibitemOpen
  \bibfield  {author} {\bibinfo {author} {\bibfnamefont {D.}~\bibnamefont
  {Paredes-Barato}}\ and\ \bibinfo {author} {\bibfnamefont {C.~S.}\
  \bibnamefont {Adams}},\ }\bibfield  {title} {\bibinfo {title} {{All-Optical
  Quantum Information Processing Using Rydberg Gates}},\ }\href
  {https://doi.org/10.1103/physrevlett.112.040501} {\bibfield  {journal}
  {\bibinfo  {journal} {Physical Review Letters}\ }\textbf {\bibinfo {volume}
  {112}},\ \bibinfo {pages} {040501} (\bibinfo {year} {2013})},\ \Eprint
  {https://arxiv.org/abs/1309.7933} {1309.7933} \BibitemShut {NoStop}%
\bibitem [{\citenamefont {Thompson}\ \emph {et~al.}(2017)\citenamefont
  {Thompson}, \citenamefont {Nicholson}, \citenamefont {Liang}, \citenamefont
  {Cantu}, \citenamefont {Venkatramani}, \citenamefont {Choi}, \citenamefont
  {Fedorov}, \citenamefont {Viscor}, \citenamefont {Pohl}, \citenamefont
  {Lukin},\ and\ \citenamefont {Vuletić}}]{Thompson.2017}%
  \BibitemOpen
  \bibfield  {author} {\bibinfo {author} {\bibfnamefont {J.~D.}\ \bibnamefont
  {Thompson}}, \bibinfo {author} {\bibfnamefont {T.~L.}\ \bibnamefont
  {Nicholson}}, \bibinfo {author} {\bibfnamefont {Q.-Y.}\ \bibnamefont
  {Liang}}, \bibinfo {author} {\bibfnamefont {S.~H.}\ \bibnamefont {Cantu}},
  \bibinfo {author} {\bibfnamefont {A.~V.}\ \bibnamefont {Venkatramani}},
  \bibinfo {author} {\bibfnamefont {S.}~\bibnamefont {Choi}}, \bibinfo {author}
  {\bibfnamefont {I.~A.}\ \bibnamefont {Fedorov}}, \bibinfo {author}
  {\bibfnamefont {D.}~\bibnamefont {Viscor}}, \bibinfo {author} {\bibfnamefont
  {T.}~\bibnamefont {Pohl}}, \bibinfo {author} {\bibfnamefont {M.~D.}\
  \bibnamefont {Lukin}},\ and\ \bibinfo {author} {\bibfnamefont
  {V.}~\bibnamefont {Vuletić}},\ }\bibfield  {title} {\bibinfo {title}
  {{Symmetry-protected collisions between strongly interacting photons}},\
  }\href {https://doi.org/10.1038/nature20823} {\bibfield  {journal} {\bibinfo
  {journal} {Nature}\ }\textbf {\bibinfo {volume} {542}},\ \bibinfo {pages}
  {206} (\bibinfo {year} {2017})}\BibitemShut {NoStop}%
\bibitem [{\citenamefont {Walker}\ and\ \citenamefont
  {Saffman}(2008)}]{Walker.2008}%
  \BibitemOpen
  \bibfield  {author} {\bibinfo {author} {\bibfnamefont {T.~G.}\ \bibnamefont
  {Walker}}\ and\ \bibinfo {author} {\bibfnamefont {M.}~\bibnamefont
  {Saffman}},\ }\bibfield  {title} {\bibinfo {title} {{Consequences of Zeeman
  degeneracy for the van der Waals blockade between Rydberg atoms}},\ }\href
  {https://doi.org/10.1103/physreva.77.032723} {\bibfield  {journal} {\bibinfo
  {journal} {Physical Review A}\ }\textbf {\bibinfo {volume} {77}},\ \bibinfo
  {pages} {032723} (\bibinfo {year} {2008})},\ \Eprint
  {https://arxiv.org/abs/0712.3438} {0712.3438} \BibitemShut {NoStop}%
\bibitem [{\citenamefont {Bohlouli-Zanjani}\ \emph {et~al.}(2007)\citenamefont
  {Bohlouli-Zanjani}, \citenamefont {Petrus},\ and\ \citenamefont
  {Martin}}]{Bohlouli-Zanjani.2007}%
  \BibitemOpen
  \bibfield  {author} {\bibinfo {author} {\bibfnamefont {P.}~\bibnamefont
  {Bohlouli-Zanjani}}, \bibinfo {author} {\bibfnamefont {J.~A.}\ \bibnamefont
  {Petrus}},\ and\ \bibinfo {author} {\bibfnamefont {J.~D.~D.}\ \bibnamefont
  {Martin}},\ }\bibfield  {title} {\bibinfo {title} {{Enhancement of Rydberg
  Atom Interactions Using ac Stark Shifts}},\ }\href
  {https://doi.org/10.1103/physrevlett.98.203005} {\bibfield  {journal}
  {\bibinfo  {journal} {Physical Review Letters}\ }\textbf {\bibinfo {volume}
  {98}},\ \bibinfo {pages} {203005} (\bibinfo {year} {2007})},\ \Eprint
  {https://arxiv.org/abs/physics/0612233} {physics/0612233} \BibitemShut
  {NoStop}%
\bibitem [{\citenamefont {Tretyakov}\ \emph {et~al.}(2014)\citenamefont
  {Tretyakov}, \citenamefont {Entin}, \citenamefont {Yakshina}, \citenamefont
  {Beterov}, \citenamefont {Andreeva},\ and\ \citenamefont
  {Ryabtsev}}]{Tretyakov.2014}%
  \BibitemOpen
  \bibfield  {author} {\bibinfo {author} {\bibfnamefont {D.~B.}\ \bibnamefont
  {Tretyakov}}, \bibinfo {author} {\bibfnamefont {V.~M.}\ \bibnamefont
  {Entin}}, \bibinfo {author} {\bibfnamefont {E.~A.}\ \bibnamefont {Yakshina}},
  \bibinfo {author} {\bibfnamefont {I.~I.}\ \bibnamefont {Beterov}}, \bibinfo
  {author} {\bibfnamefont {C.}~\bibnamefont {Andreeva}},\ and\ \bibinfo
  {author} {\bibfnamefont {I.~I.}\ \bibnamefont {Ryabtsev}},\ }\bibfield
  {title} {\bibinfo {title} {{Controlling the interactions of a few cold Rb
  Rydberg atoms by radio-frequency-assisted Förster resonances}},\ }\href
  {https://doi.org/10.1103/physreva.90.041403} {\bibfield  {journal} {\bibinfo
  {journal} {Physical Review A}\ }\textbf {\bibinfo {volume} {90}},\ \bibinfo
  {pages} {041403} (\bibinfo {year} {2014})},\ \Eprint
  {https://arxiv.org/abs/1404.0438} {1404.0438} \BibitemShut {NoStop}%
\bibitem [{\citenamefont {Ryabtsev}\ \emph {et~al.}(2010)\citenamefont
  {Ryabtsev}, \citenamefont {Tretyakov}, \citenamefont {Beterov},\ and\
  \citenamefont {Entin}}]{Ryabtsev.2010}%
  \BibitemOpen
  \bibfield  {author} {\bibinfo {author} {\bibfnamefont {I.~I.}\ \bibnamefont
  {Ryabtsev}}, \bibinfo {author} {\bibfnamefont {D.~B.}\ \bibnamefont
  {Tretyakov}}, \bibinfo {author} {\bibfnamefont {I.~I.}\ \bibnamefont
  {Beterov}},\ and\ \bibinfo {author} {\bibfnamefont {V.~M.}\ \bibnamefont
  {Entin}},\ }\bibfield  {title} {\bibinfo {title} {{Observation of the
  Stark-Tuned Förster Resonance between Two Rydberg Atoms}},\ }\href
  {https://doi.org/10.1103/physrevlett.104.073003} {\bibfield  {journal}
  {\bibinfo  {journal} {Physical Review Letters}\ }\textbf {\bibinfo {volume}
  {104}},\ \bibinfo {pages} {073003} (\bibinfo {year} {2010})},\ \Eprint
  {https://arxiv.org/abs/0909.3239} {0909.3239} \BibitemShut {NoStop}%
\bibitem [{\citenamefont {Ravets}\ \emph {et~al.}(2014)\citenamefont {Ravets},
  \citenamefont {Labuhn}, \citenamefont {Barredo}, \citenamefont {Béguin},
  \citenamefont {Lahaye},\ and\ \citenamefont {Browaeys}}]{Ravets.2014}%
  \BibitemOpen
  \bibfield  {author} {\bibinfo {author} {\bibfnamefont {S.}~\bibnamefont
  {Ravets}}, \bibinfo {author} {\bibfnamefont {H.}~\bibnamefont {Labuhn}},
  \bibinfo {author} {\bibfnamefont {D.}~\bibnamefont {Barredo}}, \bibinfo
  {author} {\bibfnamefont {L.}~\bibnamefont {Béguin}}, \bibinfo {author}
  {\bibfnamefont {T.}~\bibnamefont {Lahaye}},\ and\ \bibinfo {author}
  {\bibfnamefont {A.}~\bibnamefont {Browaeys}},\ }\bibfield  {title} {\bibinfo
  {title} {{Coherent dipole–dipole coupling between two single Rydberg atoms
  at an electrically-tuned Förster resonance}},\ }\href
  {https://doi.org/10.1038/nphys3119} {\bibfield  {journal} {\bibinfo
  {journal} {Nature Physics}\ }\textbf {\bibinfo {volume} {10}},\ \bibinfo
  {pages} {914} (\bibinfo {year} {2014})},\ \Eprint
  {https://arxiv.org/abs/1405.7804} {1405.7804} \BibitemShut {NoStop}%
\bibitem [{\citenamefont {Reinhard}\ \emph {et~al.}(2008)\citenamefont
  {Reinhard}, \citenamefont {Younge},\ and\ \citenamefont
  {Raithel}}]{Reinhard.2008}%
  \BibitemOpen
  \bibfield  {author} {\bibinfo {author} {\bibfnamefont {A.}~\bibnamefont
  {Reinhard}}, \bibinfo {author} {\bibfnamefont {K.~C.}\ \bibnamefont
  {Younge}},\ and\ \bibinfo {author} {\bibfnamefont {G.}~\bibnamefont
  {Raithel}},\ }\bibfield  {title} {\bibinfo {title} {{Effect of Förster
  resonances on the excitation statistics of many-body Rydberg systems}},\
  }\href {https://doi.org/10.1103/physreva.78.060702} {\bibfield  {journal}
  {\bibinfo  {journal} {Physical Review A}\ }\textbf {\bibinfo {volume} {78}},\
  \bibinfo {pages} {060702} (\bibinfo {year} {2008})}\BibitemShut {NoStop}%
\bibitem [{\citenamefont {Nipper}\ \emph {et~al.}(2012)\citenamefont {Nipper},
  \citenamefont {Balewski}, \citenamefont {Krupp}, \citenamefont {Hofferberth},
  \citenamefont {Löw},\ and\ \citenamefont {Pfau}}]{Nipper.2012}%
  \BibitemOpen
  \bibfield  {author} {\bibinfo {author} {\bibfnamefont {J.}~\bibnamefont
  {Nipper}}, \bibinfo {author} {\bibfnamefont {J.~B.}\ \bibnamefont
  {Balewski}}, \bibinfo {author} {\bibfnamefont {A.~T.}\ \bibnamefont {Krupp}},
  \bibinfo {author} {\bibfnamefont {S.}~\bibnamefont {Hofferberth}}, \bibinfo
  {author} {\bibfnamefont {R.}~\bibnamefont {Löw}},\ and\ \bibinfo {author}
  {\bibfnamefont {T.}~\bibnamefont {Pfau}},\ }\bibfield  {title} {\bibinfo
  {title} {{Atomic Pair-State Interferometer: Controlling and Measuring an
  Interaction-Induced Phase Shift in Rydberg-Atom Pairs}},\ }\href
  {https://doi.org/10.1103/physrevx.2.031011} {\bibfield  {journal} {\bibinfo
  {journal} {Physical Review X}\ }\textbf {\bibinfo {volume} {2}},\ \bibinfo
  {pages} {031011} (\bibinfo {year} {2012})}\BibitemShut {NoStop}%
\bibitem [{\citenamefont {Gorniaczyk}\ \emph {et~al.}(2016)\citenamefont
  {Gorniaczyk}, \citenamefont {Tresp}, \citenamefont {Bienias}, \citenamefont
  {Paris-Mandoki}, \citenamefont {Li}, \citenamefont {Mirgorodskiy},
  \citenamefont {Büchler}, \citenamefont {Lesanovsky},\ and\ \citenamefont
  {Hofferberth}}]{Gorniaczyk.2016}%
  \BibitemOpen
  \bibfield  {author} {\bibinfo {author} {\bibfnamefont {H.}~\bibnamefont
  {Gorniaczyk}}, \bibinfo {author} {\bibfnamefont {C.}~\bibnamefont {Tresp}},
  \bibinfo {author} {\bibfnamefont {P.}~\bibnamefont {Bienias}}, \bibinfo
  {author} {\bibfnamefont {A.}~\bibnamefont {Paris-Mandoki}}, \bibinfo {author}
  {\bibfnamefont {W.}~\bibnamefont {Li}}, \bibinfo {author} {\bibfnamefont
  {I.}~\bibnamefont {Mirgorodskiy}}, \bibinfo {author} {\bibfnamefont {H.~P.}\
  \bibnamefont {Büchler}}, \bibinfo {author} {\bibfnamefont {I.}~\bibnamefont
  {Lesanovsky}},\ and\ \bibinfo {author} {\bibfnamefont {S.}~\bibnamefont
  {Hofferberth}},\ }\bibfield  {title} {\bibinfo {title} {{Enhancement of
  Rydberg-mediated single-photon nonlinearities by electrically tuned Förster
  resonances}},\ }\href {https://doi.org/10.1038/ncomms12480} {\bibfield
  {journal} {\bibinfo  {journal} {Nature Communications}\ }\textbf {\bibinfo
  {volume} {7}},\ \bibinfo {pages} {12480} (\bibinfo {year} {2016})},\ \Eprint
  {https://arxiv.org/abs/1511.09445} {1511.09445} \BibitemShut {NoStop}%
\bibitem [{\citenamefont {Tanasittikosol}\ \emph {et~al.}(2011)\citenamefont
  {Tanasittikosol}, \citenamefont {Pritchard}, \citenamefont {Maxwell},
  \citenamefont {Gauguet}, \citenamefont {Weatherill}, \citenamefont
  {Potvliege},\ and\ \citenamefont {Adams}}]{Tanasittikosol.2011}%
  \BibitemOpen
  \bibfield  {author} {\bibinfo {author} {\bibfnamefont {M.}~\bibnamefont
  {Tanasittikosol}}, \bibinfo {author} {\bibfnamefont {J.~D.}\ \bibnamefont
  {Pritchard}}, \bibinfo {author} {\bibfnamefont {D.}~\bibnamefont {Maxwell}},
  \bibinfo {author} {\bibfnamefont {A.}~\bibnamefont {Gauguet}}, \bibinfo
  {author} {\bibfnamefont {K.~J.}\ \bibnamefont {Weatherill}}, \bibinfo
  {author} {\bibfnamefont {R.~M.}\ \bibnamefont {Potvliege}},\ and\ \bibinfo
  {author} {\bibfnamefont {C.~S.}\ \bibnamefont {Adams}},\ }\bibfield  {title}
  {\bibinfo {title} {{Microwave dressing of Rydberg dark states}},\ }\href
  {https://doi.org/10.1088/0953-4075/44/18/184020} {\bibfield  {journal}
  {\bibinfo  {journal} {Journal of Physics B: Atomic, Molecular and Optical
  Physics}\ }\textbf {\bibinfo {volume} {44}},\ \bibinfo {pages} {184020}
  (\bibinfo {year} {2011})},\ \Eprint {https://arxiv.org/abs/1102.0226}
  {1102.0226} \BibitemShut {NoStop}%
\bibitem [{\citenamefont {Brekke}\ \emph {et~al.}(2012)\citenamefont {Brekke},
  \citenamefont {Day},\ and\ \citenamefont {Walker}}]{Brekke.2012}%
  \BibitemOpen
  \bibfield  {author} {\bibinfo {author} {\bibfnamefont {E.}~\bibnamefont
  {Brekke}}, \bibinfo {author} {\bibfnamefont {J.~O.}\ \bibnamefont {Day}},\
  and\ \bibinfo {author} {\bibfnamefont {T.~G.}\ \bibnamefont {Walker}},\
  }\bibfield  {title} {\bibinfo {title} {{Excitation suppression due to
  interactions between microwave-dressed Rydberg atoms}},\ }\href
  {https://doi.org/10.1103/physreva.86.033406} {\bibfield  {journal} {\bibinfo
  {journal} {Physical Review A}\ }\textbf {\bibinfo {volume} {86}},\ \bibinfo
  {pages} {033406} (\bibinfo {year} {2012})},\ \Eprint
  {https://arxiv.org/abs/1207.3006} {1207.3006} \BibitemShut {NoStop}%
\bibitem [{\citenamefont {Xu}\ \emph {et~al.}(2024)\citenamefont {Xu},
  \citenamefont {Ye}, \citenamefont {Chang}, \citenamefont {Shi},\ and\
  \citenamefont {Li}}]{Xu.2024}%
  \BibitemOpen
  \bibfield  {author} {\bibinfo {author} {\bibfnamefont {B.}~\bibnamefont
  {Xu}}, \bibinfo {author} {\bibfnamefont {G.-S.}\ \bibnamefont {Ye}}, \bibinfo
  {author} {\bibfnamefont {Y.}~\bibnamefont {Chang}}, \bibinfo {author}
  {\bibfnamefont {T.}~\bibnamefont {Shi}},\ and\ \bibinfo {author}
  {\bibfnamefont {L.}~\bibnamefont {Li}},\ }\bibfield  {title} {\bibinfo
  {title} {{Continuously tunable single-photon level nonlinearity with Rydberg
  state wave-function engineering}},\ }\href
  {https://doi.org/10.1088/1361-6633/ad847e} {\bibfield  {journal} {\bibinfo
  {journal} {Reports on Progress in Physics}\ }\textbf {\bibinfo {volume}
  {87}},\ \bibinfo {pages} {110502} (\bibinfo {year} {2024})}\BibitemShut
  {NoStop}%
\bibitem [{\citenamefont {Young}\ \emph {et~al.}(2021)\citenamefont {Young},
  \citenamefont {Bienias}, \citenamefont {Belyansky}, \citenamefont {Kaufman},\
  and\ \citenamefont {Gorshkov}}]{Young.2021}%
  \BibitemOpen
  \bibfield  {author} {\bibinfo {author} {\bibfnamefont {J.~T.}\ \bibnamefont
  {Young}}, \bibinfo {author} {\bibfnamefont {P.}~\bibnamefont {Bienias}},
  \bibinfo {author} {\bibfnamefont {R.}~\bibnamefont {Belyansky}}, \bibinfo
  {author} {\bibfnamefont {A.~M.}\ \bibnamefont {Kaufman}},\ and\ \bibinfo
  {author} {\bibfnamefont {A.~V.}\ \bibnamefont {Gorshkov}},\ }\bibfield
  {title} {\bibinfo {title} {{Asymmetric Blockade and Multiqubit Gates via
  Dipole-Dipole Interactions}},\ }\href
  {https://doi.org/10.1103/physrevlett.127.120501} {\bibfield  {journal}
  {\bibinfo  {journal} {Physical Review Letters}\ }\textbf {\bibinfo {volume}
  {127}},\ \bibinfo {pages} {120501} (\bibinfo {year} {2021})},\ \Eprint
  {https://arxiv.org/abs/2006.02486} {2006.02486} \BibitemShut {NoStop}%
\bibitem [{\citenamefont {Shi}\ and\ \citenamefont {Kennedy}(2017)}]{Shi.2017}%
  \BibitemOpen
  \bibfield  {author} {\bibinfo {author} {\bibfnamefont {X.-F.}\ \bibnamefont
  {Shi}}\ and\ \bibinfo {author} {\bibfnamefont {T.~A.~B.}\ \bibnamefont
  {Kennedy}},\ }\bibfield  {title} {\bibinfo {title} {{Annulled van der Waals
  interaction and fast Rydberg quantum gates}},\ }\href
  {https://doi.org/10.1103/physreva.95.043429} {\bibfield  {journal} {\bibinfo
  {journal} {Physical Review A}\ }\textbf {\bibinfo {volume} {95}},\ \bibinfo
  {pages} {043429} (\bibinfo {year} {2017})},\ \Eprint
  {https://arxiv.org/abs/1606.08516} {1606.08516} \BibitemShut {NoStop}%
\bibitem [{\citenamefont {Shi}\ \emph {et~al.}(2016)\citenamefont {Shi},
  \citenamefont {Svetlichnyy},\ and\ \citenamefont {Kennedy}}]{Shi.2016}%
  \BibitemOpen
  \bibfield  {author} {\bibinfo {author} {\bibfnamefont {X.-F.}\ \bibnamefont
  {Shi}}, \bibinfo {author} {\bibfnamefont {P.}~\bibnamefont {Svetlichnyy}},\
  and\ \bibinfo {author} {\bibfnamefont {T.~A.~B.}\ \bibnamefont {Kennedy}},\
  }\bibfield  {title} {\bibinfo {title} {{Spin–charge separation of
  dark-state polaritons in a Rydberg medium}},\ }\href
  {https://doi.org/10.1088/0953-4075/49/7/074005} {\bibfield  {journal}
  {\bibinfo  {journal} {Journal of Physics B: Atomic, Molecular and Optical
  Physics}\ }\textbf {\bibinfo {volume} {49}},\ \bibinfo {pages} {074005}
  (\bibinfo {year} {2016})}\BibitemShut {NoStop}%
\bibitem [{\citenamefont {Sevinçli}\ and\ \citenamefont
  {Pohl}(2014)}]{Sevincli.2014}%
  \BibitemOpen
  \bibfield  {author} {\bibinfo {author} {\bibfnamefont {S.}~\bibnamefont
  {Sevinçli}}\ and\ \bibinfo {author} {\bibfnamefont {T.}~\bibnamefont
  {Pohl}},\ }\bibfield  {title} {\bibinfo {title} {{Microwave control of
  Rydberg atom interactions}},\ }\href
  {https://doi.org/10.1088/1367-2630/16/12/123036} {\bibfield  {journal}
  {\bibinfo  {journal} {New Journal of Physics}\ }\textbf {\bibinfo {volume}
  {16}},\ \bibinfo {pages} {123036} (\bibinfo {year} {2014})},\ \Eprint
  {https://arxiv.org/abs/1412.4925} {1412.4925} \BibitemShut {NoStop}%
\bibitem [{Sup()}]{Supplemental}%
  \BibitemOpen
  \href@noop {} {}\bibinfo {note} {See Supplemental Material for more information
  on the cloud shaping and microwave purification techniques, further results
  from the Floquet simulations, details of the theory model, and the
  investigation of alternative mechanisms for the reduction of
  $g^{(2)}(0)$.}\BibitemShut {Stop}%
\bibitem [{\citenamefont {Fan}\ \emph {et~al.}(2023)\citenamefont {Fan},
  \citenamefont {Zhang}, \citenamefont {Jiao}, \citenamefont {Li},
  \citenamefont {Bai}, \citenamefont {Wu}, \citenamefont {Zhao},\ and\
  \citenamefont {Jia}}]{Fan.2023}%
  \BibitemOpen
  \bibfield  {author} {\bibinfo {author} {\bibfnamefont {J.}~\bibnamefont
  {Fan}}, \bibinfo {author} {\bibfnamefont {H.}~\bibnamefont {Zhang}}, \bibinfo
  {author} {\bibfnamefont {Y.}~\bibnamefont {Jiao}}, \bibinfo {author}
  {\bibfnamefont {C.}~\bibnamefont {Li}}, \bibinfo {author} {\bibfnamefont
  {J.}~\bibnamefont {Bai}}, \bibinfo {author} {\bibfnamefont {J.}~\bibnamefont
  {Wu}}, \bibinfo {author} {\bibfnamefont {J.}~\bibnamefont {Zhao}},\ and\
  \bibinfo {author} {\bibfnamefont {S.}~\bibnamefont {Jia}},\ }\bibfield
  {title} {\bibinfo {title} {{Manipulation of single stored-photon with
  microwave field based on Rydberg polariton}},\ }\href
  {https://doi.org/10.1364/oe.487471} {\bibfield  {journal} {\bibinfo
  {journal} {Optics Express}\ }\textbf {\bibinfo {volume} {31}},\ \bibinfo
  {pages} {20641} (\bibinfo {year} {2023})}\BibitemShut {NoStop}%
\bibitem [{\citenamefont {Lampen}\ \emph {et~al.}(2018)\citenamefont {Lampen},
  \citenamefont {Nguyen}, \citenamefont {Li}, \citenamefont {Berman},\ and\
  \citenamefont {Kuzmich}}]{Lampen.2018}%
  \BibitemOpen
  \bibfield  {author} {\bibinfo {author} {\bibfnamefont {J.}~\bibnamefont
  {Lampen}}, \bibinfo {author} {\bibfnamefont {H.}~\bibnamefont {Nguyen}},
  \bibinfo {author} {\bibfnamefont {L.}~\bibnamefont {Li}}, \bibinfo {author}
  {\bibfnamefont {P.~R.}\ \bibnamefont {Berman}},\ and\ \bibinfo {author}
  {\bibfnamefont {A.}~\bibnamefont {Kuzmich}},\ }\bibfield  {title} {\bibinfo
  {title} {{Long-lived coherence between ground and Rydberg levels in a
  magic-wavelength lattice}},\ }\href
  {https://doi.org/10.1103/physreva.98.033411} {\bibfield  {journal} {\bibinfo
  {journal} {Physical Review A}\ }\textbf {\bibinfo {volume} {98}},\ \bibinfo
  {pages} {033411} (\bibinfo {year} {2018})}\BibitemShut {NoStop}%
\bibitem [{\citenamefont {Rosi}\ \emph {et~al.}(2018)\citenamefont {Rosi},
  \citenamefont {Burchianti}, \citenamefont {Conclave}, \citenamefont {Naik},
  \citenamefont {Roati}, \citenamefont {Fort},\ and\ \citenamefont
  {Minardi}}]{Rosi.2018}%
  \BibitemOpen
  \bibfield  {author} {\bibinfo {author} {\bibfnamefont {S.}~\bibnamefont
  {Rosi}}, \bibinfo {author} {\bibfnamefont {A.}~\bibnamefont {Burchianti}},
  \bibinfo {author} {\bibfnamefont {S.}~\bibnamefont {Conclave}}, \bibinfo
  {author} {\bibfnamefont {D.~S.}\ \bibnamefont {Naik}}, \bibinfo {author}
  {\bibfnamefont {G.}~\bibnamefont {Roati}}, \bibinfo {author} {\bibfnamefont
  {C.}~\bibnamefont {Fort}},\ and\ \bibinfo {author} {\bibfnamefont
  {F.}~\bibnamefont {Minardi}},\ }\bibfield  {title} {\bibinfo {title}
  {{$\Lambda$-enhanced grey molasses on the D2 transition of Rubidium-87
  atoms}},\ }\href {https://doi.org/10.1038/s41598-018-19814-z} {\bibfield
  {journal} {\bibinfo  {journal} {Scientific Reports}\ }\textbf {\bibinfo
  {volume} {8}},\ \bibinfo {pages} {1301} (\bibinfo {year} {2018})},\ \Eprint
  {https://arxiv.org/abs/1709.06467} {1709.06467} \BibitemShut {NoStop}%
\bibitem [{\citenamefont {Goldschmidt}\ \emph {et~al.}(2016)\citenamefont
  {Goldschmidt}, \citenamefont {Boulier}, \citenamefont {Brown}, \citenamefont
  {Koller}, \citenamefont {Young}, \citenamefont {Gorshkov}, \citenamefont
  {Rolston},\ and\ \citenamefont {Porto}}]{Goldschmidt.2016}%
  \BibitemOpen
  \bibfield  {author} {\bibinfo {author} {\bibfnamefont {E.~A.}\ \bibnamefont
  {Goldschmidt}}, \bibinfo {author} {\bibfnamefont {T.}~\bibnamefont
  {Boulier}}, \bibinfo {author} {\bibfnamefont {R.~C.}\ \bibnamefont {Brown}},
  \bibinfo {author} {\bibfnamefont {S.~B.}\ \bibnamefont {Koller}}, \bibinfo
  {author} {\bibfnamefont {J.~T.}\ \bibnamefont {Young}}, \bibinfo {author}
  {\bibfnamefont {A.~V.}\ \bibnamefont {Gorshkov}}, \bibinfo {author}
  {\bibfnamefont {S.~L.}\ \bibnamefont {Rolston}},\ and\ \bibinfo {author}
  {\bibfnamefont {J.~V.}\ \bibnamefont {Porto}},\ }\bibfield  {title} {\bibinfo
  {title} {{Anomalous Broadening in Driven Dissipative Rydberg Systems}},\
  }\href {https://doi.org/10.1103/physrevlett.116.113001} {\bibfield  {journal}
  {\bibinfo  {journal} {Physical Review Letters}\ }\textbf {\bibinfo {volume}
  {116}},\ \bibinfo {pages} {113001} (\bibinfo {year} {2016})},\ \Eprint
  {https://arxiv.org/abs/1510.08710} {1510.08710} \BibitemShut {NoStop}%
\bibitem [{\citenamefont {Stevens}(2013)}]{Migdall.2013}%
  \BibitemOpen
  \bibfield  {author} {\bibinfo {author} {\bibfnamefont {M.~J.}\ \bibnamefont
  {Stevens}},\ }\bibfield  {title} {\bibinfo {title} {Chapter 2 - photon
  statistics, measurements, and measurements tools},\ }in\ \href
  {https://doi.org/https://doi.org/10.1016/B978-0-12-387695-9.00002-0} {\emph
  {\bibinfo {booktitle} {Single-Photon Generation and Detection}}},\ \bibinfo
  {series} {Experimental Methods in the Physical Sciences}, Vol.~\bibinfo
  {volume} {45},\ \bibinfo {editor} {edited by\ \bibinfo {editor}
  {\bibfnamefont {A.}~\bibnamefont {Migdall}}, \bibinfo {editor} {\bibfnamefont
  {S.~V.}\ \bibnamefont {Polyakov}}, \bibinfo {editor} {\bibfnamefont
  {J.}~\bibnamefont {Fan}},\ and\ \bibinfo {editor} {\bibfnamefont {J.~C.}\
  \bibnamefont {Bienfang}}}\ (\bibinfo  {publisher} {Academic Press},\ \bibinfo
  {year} {2013})\ pp.\ \bibinfo {pages} {25--68}\BibitemShut {NoStop}%
\bibitem [{\citenamefont {Banner}\ \emph {et~al.}(2024)\citenamefont {Banner},
  \citenamefont {Kurdak}, \citenamefont {Li}, \citenamefont {Migdall},
  \citenamefont {Porto},\ and\ \citenamefont {Rolston}}]{Banner.2024}%
  \BibitemOpen
  \bibfield  {author} {\bibinfo {author} {\bibfnamefont {P.}~\bibnamefont
  {Banner}}, \bibinfo {author} {\bibfnamefont {D.}~\bibnamefont {Kurdak}},
  \bibinfo {author} {\bibfnamefont {Y.}~\bibnamefont {Li}}, \bibinfo {author}
  {\bibfnamefont {A.}~\bibnamefont {Migdall}}, \bibinfo {author} {\bibfnamefont
  {J.}~\bibnamefont {Porto}},\ and\ \bibinfo {author} {\bibfnamefont
  {S.}~\bibnamefont {Rolston}},\ }\bibfield  {title} {\bibinfo {title}
  {{Number-State Reconstruction with a Single Single-Photon Avalanche
  Detector}},\ }\bibfield  {journal} {\bibinfo  {journal} {Optica Quantum}\
  }\href {https://doi.org/10.1364/opticaq.504308} {10.1364/opticaq.504308}
  (\bibinfo {year} {2024})\BibitemShut {NoStop}%
\bibitem [{\citenamefont {Reinhard}\ \emph {et~al.}(2007)\citenamefont
  {Reinhard}, \citenamefont {Liebisch}, \citenamefont {Knuffman},\ and\
  \citenamefont {Raithel}}]{Reinhard.2007}%
  \BibitemOpen
  \bibfield  {author} {\bibinfo {author} {\bibfnamefont {A.}~\bibnamefont
  {Reinhard}}, \bibinfo {author} {\bibfnamefont {T.~C.}\ \bibnamefont
  {Liebisch}}, \bibinfo {author} {\bibfnamefont {B.}~\bibnamefont {Knuffman}},\
  and\ \bibinfo {author} {\bibfnamefont {G.}~\bibnamefont {Raithel}},\
  }\bibfield  {title} {\bibinfo {title} {{Level shifts of rubidium Rydberg
  states due to binary interactions}},\ }\href
  {https://doi.org/10.1103/physreva.75.032712} {\bibfield  {journal} {\bibinfo
  {journal} {Physical Review A}\ }\textbf {\bibinfo {volume} {75}},\ \bibinfo
  {pages} {032712} (\bibinfo {year} {2007})}\BibitemShut {NoStop}%
\bibitem [{\citenamefont {Berman}\ and\ \citenamefont
  {Kuzmich}(2024)}]{Berman.2024}%
  \BibitemOpen
  \bibfield  {author} {\bibinfo {author} {\bibfnamefont {P.~R.}\ \bibnamefont
  {Berman}}\ and\ \bibinfo {author} {\bibfnamefont {A.}~\bibnamefont
  {Kuzmich}},\ }\bibfield  {title} {\bibinfo {title} {{Interplay of the dipole
  blockade and interaction-induced dephasing in Rydberg single-photon
  sources}},\ }\href {https://doi.org/10.1103/physreva.109.013710} {\bibfield
  {journal} {\bibinfo  {journal} {Physical Review A}\ }\textbf {\bibinfo
  {volume} {109}},\ \bibinfo {pages} {013710} (\bibinfo {year}
  {2024})}\BibitemShut {NoStop}%
\bibitem [{\citenamefont {Bariani}\ \emph {et~al.}(2012)\citenamefont
  {Bariani}, \citenamefont {Goldbart},\ and\ \citenamefont
  {Kennedy}}]{Bariani.2012v2}%
  \BibitemOpen
  \bibfield  {author} {\bibinfo {author} {\bibfnamefont {F.}~\bibnamefont
  {Bariani}}, \bibinfo {author} {\bibfnamefont {P.~M.}\ \bibnamefont
  {Goldbart}},\ and\ \bibinfo {author} {\bibfnamefont {T.~A.~B.}\ \bibnamefont
  {Kennedy}},\ }\bibfield  {title} {\bibinfo {title} {{Dephasing dynamics of
  Rydberg atom spin waves}},\ }\href
  {https://doi.org/10.1103/physreva.86.041802} {\bibfield  {journal} {\bibinfo
  {journal} {Physical Review A}\ }\textbf {\bibinfo {volume} {86}},\ \bibinfo
  {pages} {041802} (\bibinfo {year} {2012})},\ \Eprint
  {https://arxiv.org/abs/1208.0355} {1208.0355} \BibitemShut {NoStop}%
\end{thebibliography}

\begin{thebibliography}{10}%
\makeatletter
\providecommand \@ifxundefined [1]{%
 \@ifx{#1\undefined}
}%
\providecommand \@ifnum [1]{%
 \ifnum #1\expandafter \@firstoftwo
 \else \expandafter \@secondoftwo
 \fi
}%
\providecommand \@ifx [1]{%
 \ifx #1\expandafter \@firstoftwo
 \else \expandafter \@secondoftwo
 \fi
}%
\providecommand \natexlab [1]{#1}%
\providecommand \enquote  [1]{``#1''}%
\providecommand \bibnamefont  [1]{#1}%
\providecommand \bibfnamefont [1]{#1}%
\providecommand \citenamefont [1]{#1}%
\providecommand \href@noop [0]{\@secondoftwo}%
\providecommand \href [0]{\begingroup \@sanitize@url \@href}%
\providecommand \@href[1]{\@@startlink{#1}\@@href}%
\providecommand \@@href[1]{\endgroup#1\@@endlink}%
\providecommand \@sanitize@url [0]{\catcode `\\12\catcode `\$12\catcode
  `\&12\catcode `\#12\catcode `\^12\catcode `\_12\catcode `\%12\relax}%
\providecommand \@@startlink[1]{}%
\providecommand \@@endlink[0]{}%
\providecommand \url  [0]{\begingroup\@sanitize@url \@url }%
\providecommand \@url [1]{\endgroup\@href {#1}{\urlprefix }}%
\providecommand \urlprefix  [0]{URL }%
\providecommand \Eprint [0]{\href }%
\providecommand \doibase [0]{https://doi.org/}%
\providecommand \selectlanguage [0]{\@gobble}%
\providecommand \bibinfo  [0]{\@secondoftwo}%
\providecommand \bibfield  [0]{\@secondoftwo}%
\providecommand \translation [1]{[#1]}%
\providecommand \BibitemOpen [0]{}%
\providecommand \bibitemStop [0]{}%
\providecommand \bibitemNoStop [0]{.\EOS\space}%
\providecommand \EOS [0]{\spacefactor3000\relax}%
\providecommand \BibitemShut  [1]{\csname bibitem#1\endcsname}%
\let\auto@bib@innerbib\@empty
\bibitem [{\citenamefont {Robinson}\ \emph {et~al.}(2021)\citenamefont
  {Robinson}, \citenamefont {Artusio-Glimpse}, \citenamefont {Simons},\ and\
  \citenamefont {Holloway}}]{Robinson.2021}%
  \BibitemOpen
  \bibfield  {author} {\bibinfo {author} {\bibfnamefont {A.~K.}\ \bibnamefont
  {Robinson}}, \bibinfo {author} {\bibfnamefont {A.~B.}\ \bibnamefont
  {Artusio-Glimpse}}, \bibinfo {author} {\bibfnamefont {M.~T.}\ \bibnamefont
  {Simons}},\ and\ \bibinfo {author} {\bibfnamefont {C.~L.}\ \bibnamefont
  {Holloway}},\ }\bibfield  {title} {\bibinfo {title} {{Atomic spectra in a
  six-level scheme for electromagnetically induced transparency and
  Autler-Townes splitting in Rydberg atoms}},\ }\href
  {https://doi.org/10.1103/physreva.103.023704} {\bibfield  {journal} {\bibinfo
   {journal} {Physical Review A}\ }\textbf {\bibinfo {volume} {103}},\ \bibinfo
  {pages} {023704} (\bibinfo {year} {2021})},\ \Eprint
  {https://arxiv.org/abs/2009.13612} {2009.13612} \BibitemShut {NoStop}%
\bibitem [{\citenamefont {Shirley}(1963)}]{Shirley.1963}%
  \BibitemOpen
  \bibfield  {author} {\bibinfo {author} {\bibfnamefont {J.~H.}\ \bibnamefont
  {Shirley}},\ }\bibfield  {title} {\bibinfo {title} {{Solution of the
  Schrödinger Equation with a Hamiltonian Periodic in Time}},\ }\href
  {https://doi.org/10.1103/physrev.138.b979} {\bibfield  {journal} {\bibinfo
  {journal} {Physical Review}\ }\textbf {\bibinfo {volume} {138}},\ \bibinfo
  {pages} {B979} (\bibinfo {year} {1963})}\BibitemShut {NoStop}%
\bibitem [{\citenamefont {Ravets}\ \emph {et~al.}(2014)\citenamefont {Ravets},
  \citenamefont {Labuhn}, \citenamefont {Barredo}, \citenamefont {Béguin},
  \citenamefont {Lahaye},\ and\ \citenamefont {Browaeys}}]{Ravets.2014}%
  \BibitemOpen
  \bibfield  {author} {\bibinfo {author} {\bibfnamefont {S.}~\bibnamefont
  {Ravets}}, \bibinfo {author} {\bibfnamefont {H.}~\bibnamefont {Labuhn}},
  \bibinfo {author} {\bibfnamefont {D.}~\bibnamefont {Barredo}}, \bibinfo
  {author} {\bibfnamefont {L.}~\bibnamefont {Béguin}}, \bibinfo {author}
  {\bibfnamefont {T.}~\bibnamefont {Lahaye}},\ and\ \bibinfo {author}
  {\bibfnamefont {A.}~\bibnamefont {Browaeys}},\ }\bibfield  {title} {\bibinfo
  {title} {{Coherent dipole–dipole coupling between two single Rydberg atoms
  at an electrically-tuned Förster resonance}},\ }\href
  {https://doi.org/10.1038/nphys3119} {\bibfield  {journal} {\bibinfo
  {journal} {Nature Physics}\ }\textbf {\bibinfo {volume} {10}},\ \bibinfo
  {pages} {914} (\bibinfo {year} {2014})},\ \Eprint
  {https://arxiv.org/abs/1405.7804} {1405.7804} \BibitemShut {NoStop}%
\bibitem [{\citenamefont {Sun}\ and\ \citenamefont
  {Robicheaux}(2008)}]{Sun.2008}%
  \BibitemOpen
  \bibfield  {author} {\bibinfo {author} {\bibfnamefont {B.}~\bibnamefont
  {Sun}}\ and\ \bibinfo {author} {\bibfnamefont {F.}~\bibnamefont
  {Robicheaux}},\ }\bibfield  {title} {\bibinfo {title} {{Numerical study of
  two-body correlation in a 1D lattice with perfect blockade}},\ }\href
  {https://doi.org/10.1088/1367-2630/10/4/045032} {\bibfield  {journal}
  {\bibinfo  {journal} {New Journal of Physics}\ }\textbf {\bibinfo {volume}
  {10}},\ \bibinfo {pages} {045032} (\bibinfo {year} {2008})},\ \Eprint
  {https://arxiv.org/abs/0805.0132} {0805.0132} \BibitemShut {NoStop}%
\bibitem [{\citenamefont {Dudin}\ \emph {et~al.}(2012)\citenamefont {Dudin},
  \citenamefont {Bariani},\ and\ \citenamefont {Kuzmich}}]{Dudin.2012sx}%
  \BibitemOpen
  \bibfield  {author} {\bibinfo {author} {\bibfnamefont {Y.~O.}\ \bibnamefont
  {Dudin}}, \bibinfo {author} {\bibfnamefont {F.}~\bibnamefont {Bariani}},\
  and\ \bibinfo {author} {\bibfnamefont {A.}~\bibnamefont {Kuzmich}},\
  }\bibfield  {title} {\bibinfo {title} {{Emergence of Spatial Spin-Wave
  Correlations in a Cold Atomic Gas}},\ }\href
  {https://doi.org/10.1103/physrevlett.109.133602} {\bibfield  {journal}
  {\bibinfo  {journal} {Physical Review Letters}\ }\textbf {\bibinfo {volume}
  {109}},\ \bibinfo {pages} {133602} (\bibinfo {year} {2012})},\ \Eprint
  {https://arxiv.org/abs/1205.4708} {1205.4708} \BibitemShut {NoStop}%
\bibitem [{\citenamefont {Bariani}\ \emph {et~al.}(2011)\citenamefont
  {Bariani}, \citenamefont {Dudin}, \citenamefont {Kennedy},\ and\
  \citenamefont {Kuzmich}}]{Bariani.2011}%
  \BibitemOpen
  \bibfield  {author} {\bibinfo {author} {\bibfnamefont {F.}~\bibnamefont
  {Bariani}}, \bibinfo {author} {\bibfnamefont {Y.~O.}\ \bibnamefont {Dudin}},
  \bibinfo {author} {\bibfnamefont {T.~A.~B.}\ \bibnamefont {Kennedy}},\ and\
  \bibinfo {author} {\bibfnamefont {A.}~\bibnamefont {Kuzmich}},\ }\bibfield
  {title} {\bibinfo {title} {{Dephasing of Multiparticle Rydberg Excitations
  for Fast Entanglement Generation}},\ }\href
  {https://doi.org/10.1103/physrevlett.108.030501} {\bibfield  {journal}
  {\bibinfo  {journal} {Physical Review Letters}\ }\textbf {\bibinfo {volume}
  {108}},\ \bibinfo {pages} {030501} (\bibinfo {year} {2011})},\ \Eprint
  {https://arxiv.org/abs/1107.3202} {1107.3202} \BibitemShut {NoStop}%
\bibitem [{\citenamefont {Bariani}\ and\ \citenamefont
  {Kennedy}(2012)}]{Bariani.2012}%
  \BibitemOpen
  \bibfield  {author} {\bibinfo {author} {\bibfnamefont {F.}~\bibnamefont
  {Bariani}}\ and\ \bibinfo {author} {\bibfnamefont {T.~A.~B.}\ \bibnamefont
  {Kennedy}},\ }\bibfield  {title} {\bibinfo {title} {{Retrieval of multiple
  spin waves from a weakly excited, metastable atomic ensemble}},\ }\href
  {https://doi.org/10.1103/physreva.85.033811} {\bibfield  {journal} {\bibinfo
  {journal} {Physical Review A}\ }\textbf {\bibinfo {volume} {85}},\ \bibinfo
  {pages} {033811} (\bibinfo {year} {2012})},\ \Eprint
  {https://arxiv.org/abs/1112.1735} {1112.1735} \BibitemShut {NoStop}%
\bibitem [{\citenamefont {Maxwell}\ \emph {et~al.}(2013)\citenamefont
  {Maxwell}, \citenamefont {Szwer}, \citenamefont {Paredes-Barato},
  \citenamefont {Busche}, \citenamefont {Pritchard}, \citenamefont {Gauguet},
  \citenamefont {Weatherill}, \citenamefont {Jones},\ and\ \citenamefont
  {Adams}}]{Maxwell.2013}%
  \BibitemOpen
  \bibfield  {author} {\bibinfo {author} {\bibfnamefont {D.}~\bibnamefont
  {Maxwell}}, \bibinfo {author} {\bibfnamefont {D.~J.}\ \bibnamefont {Szwer}},
  \bibinfo {author} {\bibfnamefont {D.}~\bibnamefont {Paredes-Barato}},
  \bibinfo {author} {\bibfnamefont {H.}~\bibnamefont {Busche}}, \bibinfo
  {author} {\bibfnamefont {J.~D.}\ \bibnamefont {Pritchard}}, \bibinfo {author}
  {\bibfnamefont {A.}~\bibnamefont {Gauguet}}, \bibinfo {author} {\bibfnamefont
  {K.~J.}\ \bibnamefont {Weatherill}}, \bibinfo {author} {\bibfnamefont
  {M.~P.~A.}\ \bibnamefont {Jones}},\ and\ \bibinfo {author} {\bibfnamefont
  {C.~S.}\ \bibnamefont {Adams}},\ }\bibfield  {title} {\bibinfo {title}
  {{Storage and Control of Optical Photons Using Rydberg Polaritons}},\ }\href
  {https://doi.org/10.1103/physrevlett.110.103001} {\bibfield  {journal}
  {\bibinfo  {journal} {Physical Review Letters}\ }\textbf {\bibinfo {volume}
  {110}},\ \bibinfo {pages} {103001} (\bibinfo {year} {2013})},\ \Eprint
  {https://arxiv.org/abs/1207.6007} {1207.6007} \BibitemShut {NoStop}%
\bibitem [{\citenamefont {Maxwell}\ \emph {et~al.}(2014)\citenamefont
  {Maxwell}, \citenamefont {Szwer}, \citenamefont {Paredes-Barato},
  \citenamefont {Busche}, \citenamefont {Pritchard}, \citenamefont {Gauguet},
  \citenamefont {Jones},\ and\ \citenamefont {Adams}}]{Maxwell.2014}%
  \BibitemOpen
  \bibfield  {author} {\bibinfo {author} {\bibfnamefont {D.}~\bibnamefont
  {Maxwell}}, \bibinfo {author} {\bibfnamefont {D.~J.}\ \bibnamefont {Szwer}},
  \bibinfo {author} {\bibfnamefont {D.}~\bibnamefont {Paredes-Barato}},
  \bibinfo {author} {\bibfnamefont {H.}~\bibnamefont {Busche}}, \bibinfo
  {author} {\bibfnamefont {J.~D.}\ \bibnamefont {Pritchard}}, \bibinfo {author}
  {\bibfnamefont {A.}~\bibnamefont {Gauguet}}, \bibinfo {author} {\bibfnamefont
  {M.~P.~A.}\ \bibnamefont {Jones}},\ and\ \bibinfo {author} {\bibfnamefont
  {C.~S.}\ \bibnamefont {Adams}},\ }\bibfield  {title} {\bibinfo {title}
  {{Microwave control of the interaction between two optical photons}},\ }\href
  {https://doi.org/10.1103/physreva.89.043827} {\bibfield  {journal} {\bibinfo
  {journal} {Physical Review A}\ }\textbf {\bibinfo {volume} {89}},\ \bibinfo
  {pages} {043827} (\bibinfo {year} {2014})},\ \Eprint
  {https://arxiv.org/abs/1308.1425} {1308.1425} \BibitemShut {NoStop}%
\bibitem [{\citenamefont {Xu}\ \emph {et~al.}(2024)\citenamefont {Xu},
  \citenamefont {Ye}, \citenamefont {Chang}, \citenamefont {Shi},\ and\
  \citenamefont {Li}}]{Xu.2024}%
  \BibitemOpen
  \bibfield  {author} {\bibinfo {author} {\bibfnamefont {B.}~\bibnamefont
  {Xu}}, \bibinfo {author} {\bibfnamefont {G.-S.}\ \bibnamefont {Ye}}, \bibinfo
  {author} {\bibfnamefont {Y.}~\bibnamefont {Chang}}, \bibinfo {author}
  {\bibfnamefont {T.}~\bibnamefont {Shi}},\ and\ \bibinfo {author}
  {\bibfnamefont {L.}~\bibnamefont {Li}},\ }\bibfield  {title} {\bibinfo
  {title} {{Continuously tunable single-photon level nonlinearity with Rydberg
  state wave-function engineering}},\ }\href
  {https://doi.org/10.1088/1361-6633/ad847e} {\bibfield  {journal} {\bibinfo
  {journal} {Reports on Progress in Physics}\ }\textbf {\bibinfo {volume}
  {87}},\ \bibinfo {pages} {110502} (\bibinfo {year} {2024})}\BibitemShut
  {NoStop}%
\end{thebibliography}
\end{document}